\documentclass{vgtc}         

\ifpdf
  \pdfoutput=1\relax                   
  \usepackage{graphicx}                
  \DeclareGraphicsExtensions{.pdf,.png,.jpg,.jpeg} 
\else
  \usepackage{graphicx}                
  \DeclareGraphicsExtensions{.eps}     
\fi%
\graphicspath{{figures/}{pictures/}{images/}{./}} 

\PassOptionsToPackage{dvipsnames}{xcolor}
\vgtcinsertpkg

\usepackage{url}
\usepackage[utf8]{inputenc}
\usepackage[T1]{fontenc}
\usepackage{microtype}                 
\PassOptionsToPackage{warn}{textcomp}  
\usepackage{textcomp}                  
\usepackage{mathptmx}                  
\usepackage{times}                     
\usepackage{cite}                      
\usepackage{booktabs}                  
\usepackage{amsmath}
\usepackage{multirow}
\usepackage{cleveref}
\usepackage{t1enc}
\usepackage{tabularx}
\usepackage{makecell}
\usepackage{xspace}
\usepackage{soul}
\usepackage{color}
\usepackage{relsize}
\usepackage[plain]{algorithm}
\usepackage[noend]{algpseudocode}
\usepackage[labelfont=sf]{subcaption}
\onlineid{1002}

\vgtccategory{Research}
\vgtcpapertype{algorithm/technique}

\title{Speculative Progressive Raycasting for Memory Constrained Isosurface Visualization of Massive Volumes}

\author{Will Usher\thanks{will@willusher.io}\\ %
    \parbox{1.4in}{\scriptsize \centering Now at Luminary Cloud \\ Intel}
\and Landon Dyken\\ %
    \scriptsize University of Illinois at Chicago %
\and Sidharth Kumar\\ %
    \scriptsize University of Illinois at Chicago}

\abstract{%
New web technologies have enabled the deployment
of powerful GPU-based computational pipelines that run entirely in the web browser,
opening a new frontier for accessible scientific visualization applications.
However, these new capabilities do not address the memory
constraints of lightweight end-user devices encountered when attempting to visualize the massive data sets
produced by today's simulations and data acquisition systems.
In this paper, we propose a novel implicit isosurface rendering algorithm for
interactive visualization of massive volumes within a small memory footprint.
We achieve this by progressively traversing a wavefront of rays through the volume and
decompressing blocks of the data on-demand to perform implicit ray-isosurface intersections.
The progressively rendered surface is displayed after each pass to improve interactivity.
Furthermore, to accelerate rendering and increase GPU utilization, we introduce speculative
ray-block intersection into our algorithm, where additional blocks are traversed and intersected
speculatively along rays as other rays terminate to exploit additional parallelism in the workload.
Our entire pipeline is run in parallel on the GPU to leverage the parallel computing power
that is available even on lightweight end-user devices.
We compare our algorithm to the state of the art in low-overhead isosurface extraction
and demonstrate that it achieves $1.7\times$--$5.7\times$ reductions in memory overhead
and up to $8.4\times$ reductions in data decompressed.
}

\teaser{
 \centering
 \vspace{-1.5em}
 \includegraphics[trim={0 5.5cm 0 4.8cm},clip,width=0.98\linewidth]{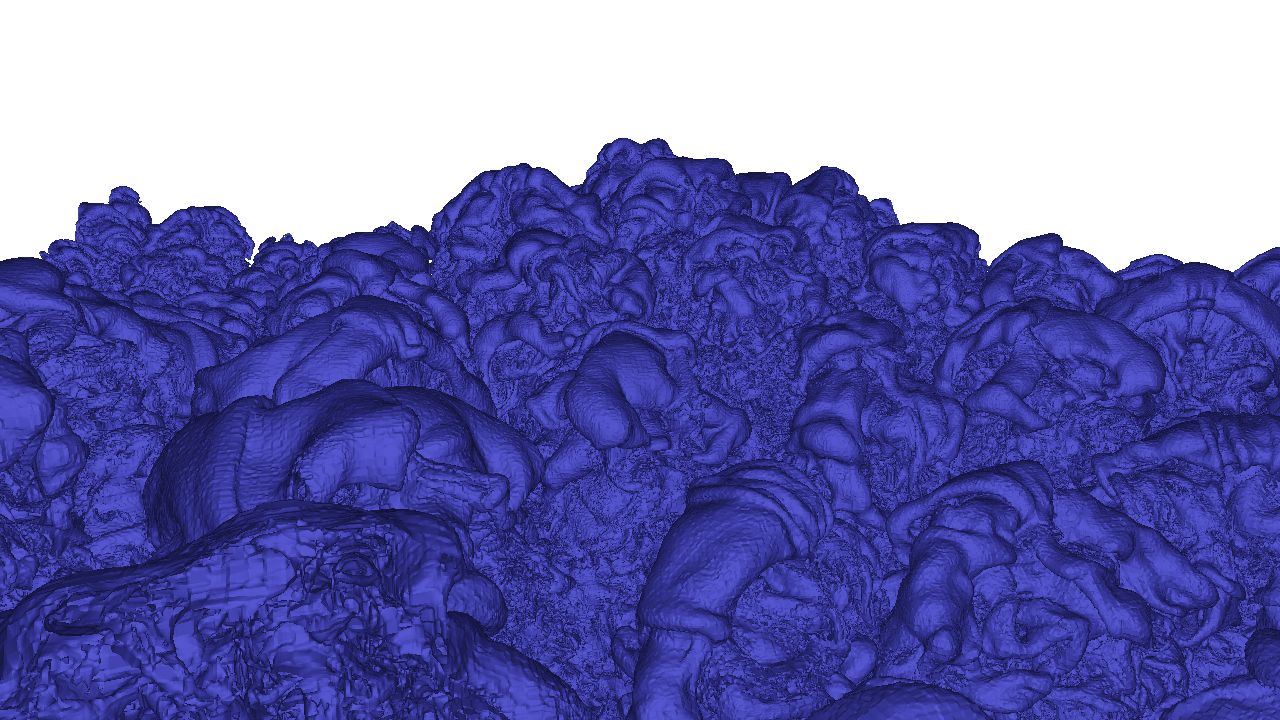}
 \vspace{-0.5em}
  \caption{\label{fig:teaser}%
  Interactive full-resolution isosurface visualization of the $2048 \times 2048 \times 1920$ Richtmyer-Meshkov (R-M) data set in the browser.
  We propose a new GPU algorithm for implicit isosurface rendering that progressively
  traverses rays through the volume and decompresses data on-demand to minimize its memory footprint.
  We achieve up to $5.7\times$ reductions in overall memory use and $8.4\times$ reductions
  in data decompressed compared to the state of the art
  in memory constrained isosurface extraction~\cite{usher_interactive_2020},
  without sacrificing interactivity.
  At $1280\times 720$, the Richtmyer-Meshkov averages 264ms per-pass and 1.2s total on an RTX~3080.}
}

\begin{document}

\firstsection{Introduction}

\maketitle

\label{sec:introduction}

Recent advances in web technologies, specifically WebGPU~\cite{webgpu} and WebAssembly~\cite{webassembly},
have enabled the development of powerful GPU-based compute applications that run
entirely in the browser. Scientific visualization applications can leverage these
technologies to take advantage of the ease of deployment afforded by the browser
without sacrificing the compute capabilities required to perform their analysis and visualization tasks;
thereby widening the accessibility of complex scientific visualization applications.
For example, recent works have leveraged these new technologies for interactive
isosurface extraction on compressed data on the GPU~\cite{usher_interactive_2020} and
GPU-parallel layout computation of large graphs~\cite{dyken_graphwagu_2022}.

However, these new technologies alone do not address the fundamental issues of
limited memory and compute capacity on lightweight end-user devices.
Memory capacity constraints are a fundamental issue in scientific
visualization even when targeting high-end workstations, and are especially
problematic when faced with processing
the massive data sets produced by current simulations and data acquisition systems
on lightweight consumer GPUs.
While there exists a large body of work
on large-scale volume rendering approaches~\cite{beyer_gpuvolstar_2015},
deploying large-scale volume visualization in the browser poses its own
unique set of additional challenges (see, e.g.~\cite{usher_interactive_2020}).
Desktop large-scale volume rendering approaches typically leverage special
purpose file formats to stream data from disk (e.g.~\cite{crassin_gigavoxels_2009,hadwiger_interactive_2012,hadwiger_sparseleap_2018,fogal_analysis_2013}); however, web applications
are unable to perform such low-level disk I/O operations.
Although prior work has leveraged remote servers to stream subsets of data~\cite{schutz_potree_2016,sherif_brainbrowser_2015},
this introduces tradeoffs with latency and deployment cost.



Usher and Pascucci~\cite{usher_interactive_2020} recently proposed Block-Compressed Marching Cubes (BCMC)
to achieve interactive isosurface extraction in the browser through on the fly decompression and caching
of a compressed data set stored on the GPU.
Their approach reduces latency by transferring the entire compressed volume to the client,
eliminating the need for a complex server,
and achieves interactive isosurface extraction times through a fully GPU-driven decompression, caching,
and isosurfacing pipeline.
However, their approach extracts explicit surface geometry and thus, as with other
extraction techniques, its memory and compute
costs scale with the size of the data set and the number of triangles in the surface.
As a result, BCMC is unable to extract isosurfaces from large data sets on
lightweight devices as it runs out of memory to store the vertex data.

We begin from the on the fly GPU decompression strategy of Usher and Pascucci~\cite{usher_interactive_2020};
however, we make deliberate design choices to reduce memory consumption
and the impact of data set size on memory footprint and compute cost.
First, we eliminate the need to store a large triangle mesh for the surface by
adopting an implicit ray-isosurface intersection approach~\cite{marmitt_fast_2004}.
Next, to avoid processing fully occluded blocks,
we progressively traverse a wavefront of rays through the volume in a multipass approach.
Volume data cache updates are performed each pass to decompress newly visible blocks,
thereby reducing the working set to just those blocks that the current set of rays
traverse in a given pass.
Finally, to address utilization issues encountered as rays terminate, we introduce
ray-block speculation into our algorithm to exploit additional parallelism on the GPU
to terminate rays faster and accelerate rendering.
Our algorithm can be easily scaled down to run on low power devices, as its costs
are primarily tied to the image size.
Our contributions are:
\begin{itemize}
    \item A novel progressive algorithm for implicit isosurface raycasting that works
        directly on compressed data on the GPU;
    \item A per-pass view-dependent decompression and caching strategy
        built into the algorithm to minimize its memory footprint;
    \item A dynamic work speculation strategy that exploits additional parallelism
        in the workload to increase GPU utilization and accelerate rendering completion;
    \item Evaluation of our algorithm against the state of the art
    on data sets with up to 8.05B voxels on lightweight end user devices.
\end{itemize}



\section{Related Work}
\label{sec:related_work}
In~\Cref{sec:related_work_browser_vis}, we review recent work on bringing scientific visualization
to the browser through WebGL~\cite{WebGL} and WebGPU~\cite{webgpu}.
Visualizing large-scale volumetric data is a fundamental problem in
scientific visualization, and has been deeply explored (see surveys by Beyer et al.~\cite{beyer_gpuvolstar_2015}
and Rodriguez et al.~\cite{balsa_rodriguez_state---art_2014-1}).
Isosurface visualization techniques can be categorized as either explicit
surface extraction methods (\Cref{sec:related_work_extract_iso}),
where geometry is computed for the surface,
or implicit surface rendering methods (\Cref{sec:related_work_implicit_iso}),
which directly compute ray-isosurface intersections without explicit geometry.
Finally, due to the similarities in isosurface ray-casting
and ray-guided volume rendering algorithms, we review relevant work on raycasting
of large volumes in~\Cref{sec:related_work_large_volume_rendering}.


\begin{figure*}
    \centering
    \vspace{-1em}
    \includegraphics[width=0.95\textwidth]{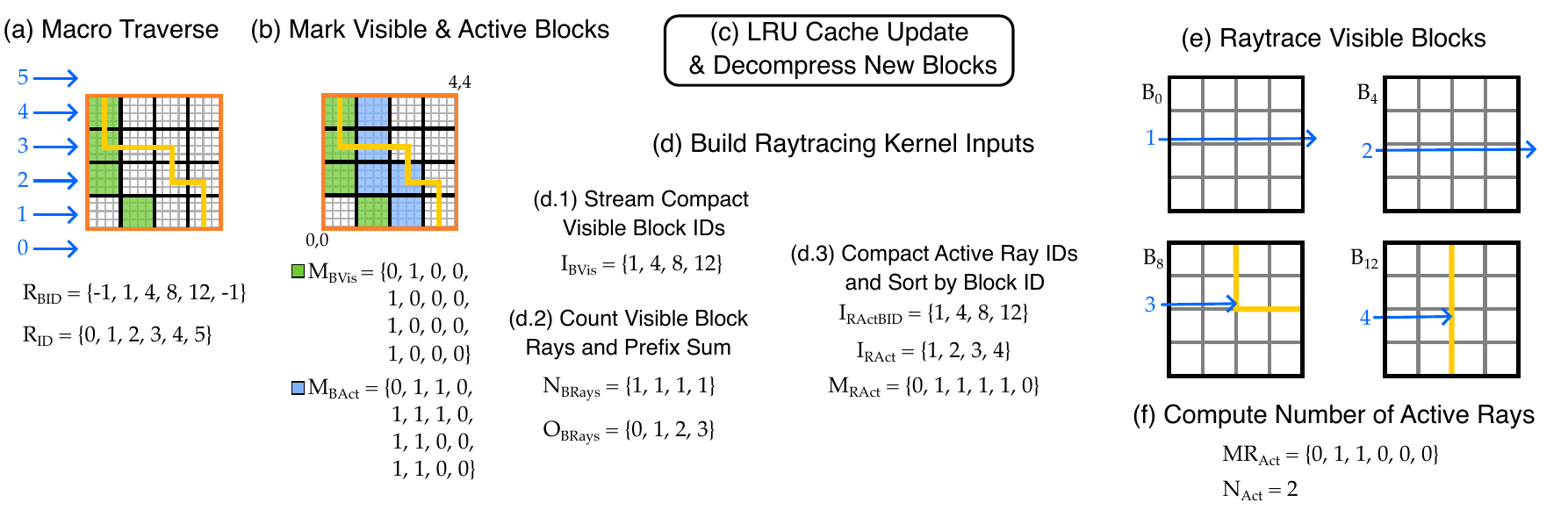}
    \vspace{-1.25em}
    \caption{\label{fig:method_pipeline_sketch}%
    An illustration of our algorithm's core loop
    on a slice of a $16^3$ volume. (a) The volume has a single coarse macrocell
    (orange) with a $4^3$ grid of ZFP blocks within it.
    After computing the initial set of rays we
    repeat steps (a-f) until all rays have terminated, displaying the partial
    image after each pass.
    (a) Rays are advanced to the next active block (green blocks), storing the block ID in $R_{BID}$.
    Rays one and two traverse blocks whose value range combined with their neighbors
    contains the isovalue, indicating that their dual grid may contain the isosurface and thus must be traversed.
    (b) The blocks referenced in $R_{BID}$ are marked visible
    and active ($M_{BVis}$, green blocks),
    and their neighbors to the positive side marked active ($M_{BAct}$, blue blocks).
    The neighbors will be required to populate the dual grid vertices for the visible blocks.
    (c) $M_{BAct}$ is passed to the LRU cache~\cite{usher_interactive_2020}, which decompresses
    and caches any new blocks required for the pass, potentially evicting those that are no longer needed.
    (d) We then prepare the inputs for the block raytracing kernel through stream compactions
    and parallel sorts on the GPU.
    (e) Each block traverses its rays through its local data, terminating those that
    intersect the isosurface.
    (f) Finally, we compute the remaining number of active rays to determine
    if rendering is complete and display the current image to the user.}
    \vspace{-1.5em}
\end{figure*}

\subsection{Scientific Visualization in the Browser}
\label{sec:related_work_browser_vis}
As with information visualization applications, bringing scientific visualization to the
browser greatly expands accessibility to visualization, enabling more scientists
to gain better insights about their data.
Prior work has brought compelling scientific visualization applications to the browser through
the use of server-side processing, local GPU acceleration, and combinations of both techniques.

Prior server-based techniques have moved all computation
to the server and streamed images to the client~\cite{perez_ipython_2007,raji_scalable_2017,raji_scientific_2018,dykes_interactive_2018,jourdain_paraviewweb_2011},
allowing lightweight clients to access large amounts of compute power.
However, such approaches can face issues with latency, cost, and quality of service when faced
with supporting large numbers of concurrent users.
Prior work has demonstrated leveraging a remote server to query and stream subsets
of data to the client~\cite{schutz_potree_2016,sherif_brainbrowser_2015}, thereby balancing
between remote and local processing costs.
Clients can query subsets of the data for their region of interest or level of detail, which is transferred
and rendered or processed locally.
Although moving the rendering work to the client reduces the impact of server latency and quality of service,
data streaming approaches can face similar issues as fully server-side approaches at scale.

In this work, we target a fully client-side processing approach to eliminate the need
for running backend servers and related potential challenges.
We note that a combination of client- and server-side processing can provide
the best scalability and performance for large data visualization; here
we focus on expanding the capabilities of the client.
Prior to WebGPU, browser applications leveraged WebGL to perform GPU accelerated 
rendering in applications ranging from LiDAR visualization~\cite{schutz_potree_2016}
to volume rendering~\cite{mobeen_high-performance_2012} and neuroscience~\cite{sherif_brainbrowser_2015,jacinto_web_2012}.
A fundamental limitation of WebGL compared to WebGPU is the lack of support for general compute shaders; however,
Li and Ma~\cite{li_p4_2018,li_p5_2019} proposed a method to work around this limitation by repurposing
the rendering pipeline to perform a subset of parallel compute operations.

With the recent development of WebGPU, browser applications now have access to
general purpose GPU compute and advanced rendering capabilities.
Usher and Pascucci~\cite{usher_interactive_2020} leveraged WebGPU
to deploy a GPU-driven isosurface extraction pipeline that achieved interactive
visualization of massive data sets entirely in the browser.
Dyken et al.~\cite{dyken_graphwagu_2022} presented a graph layout algorithm
in WebGPU to accelerate layout and rendering of large graphs.
Hidaka et al.~\cite{hidaka_dnn_2017} accelerated deep neural network execution in the browser
using WebGPU, an approach which is also being tested in TensorFlow.

\subsection{Explicit Isosurface Extraction}
\label{sec:related_work_extract_iso}
The original Marching Cubes paper~\cite{lorenson_marching_1987} defined an object-order technique that computed
explicit triangle geometry for each voxel to render the isosurface.
The extracted triangle geometry can typically be rendered in real-time on modern GPUs.
Subsequent work proposed constructing interval trees~\cite{cignoni_optimal_1996} or $k$-d trees over the span
space~\cite{livnat_near_1996} to accelerate Marching Cubes by filtering out voxels that were known to not contain
the isosurface.
Isosurface meshes can contain large numbers of triangles, many of which will be occluded or subpixel in size for a given viewpoint,
leading to wasted computation and memory use.
Livnat and Hansen~\cite{livnat_view_1998} proposed a view-dependent surface extraction technique
that traversed an octree to find voxels to extract triangles from.
Extracted triangles were rasterized for display and updated an occlusion buffer used to filter out
occluded octree nodes to skip traversal of occluded regions.


Recent work has primarily focused on leveraging parallel execution on GPUs to accelerate surface extraction~\cite{cuda_mc_example, ciznicky_efficient_2012, dyken_high-speed_2008, liu_parallel_2016, martin_load-balanced_2010, schmitz_efficient_2010, schroeder_flying_2015, usher_interactive_2020,kreskowski_efficient_2022}.
Although each voxel can be processed independently in parallel, coordination is required
to ensure that the individual voxel's outputs do not overwrite each other.
GPU-parallel algorithms achieve this through prefix sums and stream compactions to compute which
voxels contain the surface and to assign offsets into output buffers for their triangle data.
However, prior work typically assumes that the entire volume fits in the memory of a single GPU~\cite{dyken_high-speed_2008,liu_parallel_2016,schmitz_efficient_2010,schroeder_flying_2015,kreskowski_efficient_2022},
or that it can be distributed over a cluster~\cite{martin_load-balanced_2010}.

Usher and Pascucci~\cite{usher_interactive_2020} recently proposed the Block-Compressed Marching Cubes (BCMC) algorithm
for interactive GPU-parallel isosurface extraction on massive data sets.
Their approach uploads a ZFP fixed-rate compressed volume to the GPU and decompresses and caches the blocks required
for a given isosurface on demand using GPU decompression and an LRU cache.
BCMC achieves interactive isosurface extraction times on consumer GPUs;
however, as with other surface extraction techniques, it produces large vertex buffers
and its cost scales with the total number of blocks containing the isosurface.
These factors limit BCMCs scalablity to massive volumes and lightweight end-user systems.
Although we adopt a similar on-demand decompression and caching strategy,
our algorithm does not store a vertex buffer and processes blocks in a
view-dependent wavefront. These design choices significantly reduce
our algorithm's memory footprint, and tie compute costs primarily to
image size to provide better control over compute cost on lightweight devices.

\subsection{Implicit Isosurface Rendering}
\label{sec:related_work_implicit_iso}
Parker et al.~\cite{parker_interactive_1998} proposed the first implicit isosurface rendering technique,
where rays were traversed through the volume grid and ray-voxel intersections computed directly by
solving a cubic polynomial.
Parker et al.~\cite{parker_interactive_1998} accelerated ray-traversal by skipping empty space using
a multi-level grid hierarchy.
Marmitt et al.~\cite{marmitt_fast_2004} improved the quality and speed of ray-voxel intersection
through a root finding approach based on isolation and iterative root finding.
Wald et al.~\cite{wald_faster_2005} further accelerated empty space skipping through an implicit $k$-d tree
tracking value ranges of subregions of the volume.

Hadwiger et al.~\cite{hadwiger_realtime_2005} proposed an implicit isosurface rendering
technique that combined object and image order empty skipping to accelerate rendering,
coupled with a brick cache to reduce memory use.
Their algorithm constructs a fine grid over the volume and rasterizes the front and back faces of cells
that potentially contain the isosurface to generate ray start and end positions,
then performs ray marching on the GPU to find isosurface intersections between these intervals.
Hadwiger et al. employed a brick cache using a coarse grid to reduce memory use, where
data for a grid cell is only uploaded to the GPU if its value range contains the isovalue.
However, this caching strategy does not take into account visibility, and
as such will upload data for occluded regions of the volume.
In contrast, our algorithm performs data decompression on-demand as rays traverse the volume,
reducing the working set to just the blocks visible in a single pass.
Moreover, our proposed decompression and caching pipeline runs entirely on the GPU to eliminate
CPU communication bottlenecks.



\subsection{Ray-guided Large Volume Rendering}
\label{sec:related_work_large_volume_rendering}
A large body of work has explored techniques to address memory constraints in
ray-guided volume rendering~\cite{beyer_gpuvolstar_2015,crassin_gigavoxels_2009,engel_cera-tvr_2011,hadwiger_interactive_2012,hadwiger_sparseleap_2018,fogal_analysis_2013},
which we briefly review here due to their applicability to implicit isosurface raycasting.
Ray-guided techniques for large volume rendering typically combine GPU-driven cache requests,
made as rays encounter missing data during traversal,
with a CPU-side data management system that services these requests by uploading
new data to the GPU~\cite{crassin_gigavoxels_2009,hadwiger_interactive_2012,hadwiger_sparseleap_2018,fogal_analysis_2013}.
The CPU-side data management system is typically coupled with a special purpose
file format and takes advantage of low-level file system APIs to efficiently stream
massive data sets off disk.
Prior work has demonstrated interactive rendering of data sets ranging in size from
hundreds of gigabytes~\cite{fogal_analysis_2013} to terabytes~\cite{crassin_gigavoxels_2009,hadwiger_interactive_2012,hadwiger_sparseleap_2018}.

Volume rendering techniques that operate on compressed data have been proposed to alleviate disk space and in-memory working set requirements~\cite{balsa_rodriguez_state---art_2014-1,mensmann_gpu-supported_2010,suter_interactive_2011,wang_application-driven_2010,fout_high-quality_2005,schneider_compression_2003,fout_transform_2007,gobbetti_covra_2012}.
Schneider and Westermann~\cite{schneider_compression_2003} proposed a hierarchical quantization scheme that
similarly decomposes the data into $4^3$ blocks and computes a $1/4$ resolution quantized representation of each block
combined with two codebooks for the volume. Samples are then reconstructed in a slice-based renderer using
the quantized volume and codebooks. 
Fout et al.~\cite{fout_high-quality_2005,fout_transform_2007} proposed a vector quantization technique
combined with deferred filtering for slice-based rendering.
Rendering occurs in two-passes for each slice, slices are first decompressed
to a small cache, after which filtering and blending is performed.
Subsequent works have leveraged compressed GPU texture formats~\cite{mensmann_gpu-supported_2010},
combining bricking, quantization and run-length encoding~\cite{wang_application-driven_2010},
and extending these techniques with multiresolution data representations using tensor approximations~\cite{suter_interactive_2011} and octrees over compressed blocks~\cite{gobbetti_covra_2012}.

As with prior work, we adopt a brick-based compression scheme to allow decompression
of spatial subregions on-demand, leveraging ZFP~\cite{lindstrom_fixed_rate_2014}
to compress the bricks.
We note that it would be possible to leverage other brick-based compression schemes,
and to combine our technique with multiresolution level of detail hierarchies
to address undersampling issues, or out-of-core streaming methods to support larger data sets.
Finally, our choice of only using ZFP for compressing the data amounts to using
only the ``precision'' axis for data reduction; however, better data reduction
and quality can be achieved by combining the precision and resolution axes~\cite{hoang_study_2019}.

\section{Progressive Wavefront Isosurface Raycasting}
\label{sec:method}
\begin{figure*}[ht]
    \centering
    \vspace{-1em}
    \includegraphics[width=0.95\textwidth]{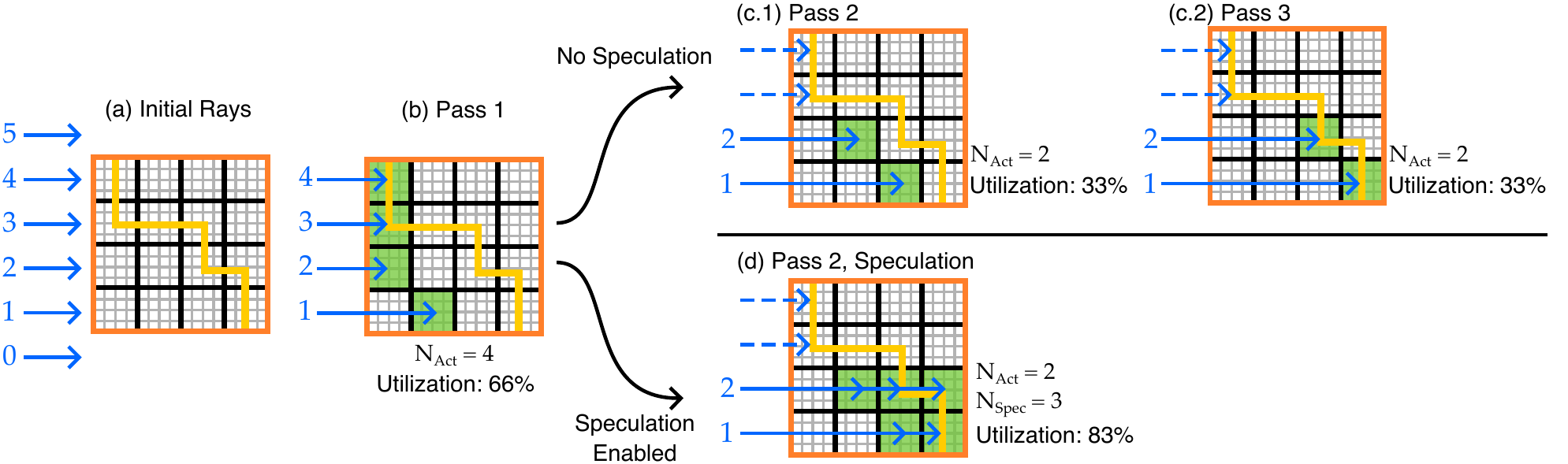}
    \vspace{-0.5em}
    \caption{\label{fig:method_traversal_speculation}%
    An illustration of the ray traversal passes for an example isosurface on a $16^3$ volume,
    without (b, c.1, c.2) and with (b, d) speculation. Green squares mark the blocks currently being traversed by a ray.
    (a) Four of the six initial rays intersect the volume's bounds.
    (b) Pass one is identical in both cases, as not enough rays have terminated to enable speculation.
    (c.1, c.2) Without speculation, rays one and two traverse one block at a time until
    they hit the isosurface, requiring two additional passes with low GPU utilization to complete the rendering.
    (d) With speculation, enough rays have terminated after pass one that $N_\text{Spec}=3$,
    increasing utilization to $83\%$ and completing the rendering in one additional pass
    by intersecting rays one and two against multiple blocks.
    A trade-off of speculative execution strategies is the potential for wasted computation.
    This is illustrated by ray two, which traverses an extra occluded block in pass two.
    Overall, our speculative execution strategy significantly reduces the total number of passes,
    and thus total time, required to render isosurfaces.}
    \vspace{-1.5em}
\end{figure*}

Our algorithm is designed with a focus on reducing overall memory consumption
and on achieving scalable and controllable rendering performance that is
not strongly impacted by the data set size.
These properties enable the algorithm to be used for visualizing massive
data sets in the browser on lightweight end user devices.
To achieve this, we propose an implicit isosurface raycasting algorithm
that progressively traverses a wavefront of rays through a block-compressed volume
(\Cref{fig:method_pipeline_sketch}).
In each pass, new visible blocks that potentially contain the isosurface are decompressed
and cached in an LRU cache to enable re-use of decompressed blocks across passes.
Thus our algorithm's memory footprint and compute cost is dependent on the image size,
view position, and isovalue.
The progressively rendered image is displayed after each pass to improve interactivity.

At a high-level, our algorithm proceeds as follows.
First, volume data compressed using
ZFP's~\cite{lindstrom_fixed_rate_2014} fixed-rate compression mode is uploaded to the GPU.
We then construct a two-level macrocell grid~\cite{parker_interactive_1998} on the GPU
to accelerate ray traversal (\Cref{sec:method_macrogrid_construction}).
For each new isovalue or camera viewpoint, we compute the view rays for each pixel (\Cref{sec:method_inital_rays}).
The following steps are then repeated to traverse the wavefront of rays
through the volume to progressively render the isosurface (see~\Cref{fig:method_pipeline_sketch}).
First, we traverse the rays through the macrocell grids to find the next block they
must test for intersections with (\Cref{fig:method_pipeline_sketch}a, \Cref{sec:method_macrogrid_traversal}).
We then mark all the blocks that are visible or active
in the current pass (\Cref{fig:method_pipeline_sketch}b, \Cref{sec:method_mark_blocks}).
The data for uncached blocks are decompressed using a WebGPU port of ZFP's CUDA
decompressor, and cached for re-use between passes through a GPU-driven LRU cache,
as done by Usher and Pascucci~\cite{usher_interactive_2020}
(\Cref{fig:method_pipeline_sketch}c, \Cref{sec:method_lru_cache}).
We then construct arrays of the visible block IDs, the number of rays intersecting each block,
and the ray IDs sorted by their block ID to provide inputs to the block raycasting kernel
(\Cref{fig:method_pipeline_sketch}d, \Cref{sec:method_compact_sort_ray_block_ids}).
Each block then intersects its rays with its local region of data to find ray-isosurface intersections (\Cref{fig:method_pipeline_sketch}e, \Cref{sec:method_rt_visible_blocks}).
Finally, we compute the remaining number of active rays to determine if rendering has completed (\Cref{fig:method_pipeline_sketch}f) and display the current image.

\subsection{Macrocell Grid Construction}
\label{sec:method_macrogrid_construction}
As done by Usher and Pascucci~\cite{usher_interactive_2020}, we leverage the $4^3$ block decomposition
of the volume used by ZFP's fixed-rate compression mode to define a macrocell grid over the volume.
The macrocell grid is used to skip blocks that do not contain the isovalue~\cite{parker_interactive_1998},
and thereby skip decompressing them.
In addition to the ZFP block macrocell grid, referred to as the fine grid, we compute a
coarse macrocell grid by grouping $4^3$ regions of ZFP blocks to form coarse cells.
Each coarse cell contains $16^3$ voxels, allowing larger regions of space to be skipped more efficiently
to accelerate rendering of sparse isosurfaces.
The value range of each cell in the fine (or coarse) grid is computed by combining the range of the cell's
voxels (or blocks) with those of its neighbors in the $+x/y/z$ direction.
The neighbor ranges are required to ensure we do not miss values contained in the cell's
dual grid, which would lead to cracks.

When a new volume is loaded, we compute the value range of each block and
then combine each cell's range with its neighbors to populate the coarse and fine grids.
These computations are run in parallel on the GPU.
We note that our approach can be combined with an octree
or other hierachical multiresolution acceleration structure over the ZFP blocks for LOD,
rather than a two-level grid.




\subsection{Compute Initial Rays}
\label{sec:method_inital_rays}

For each new camera position or isovalue, we begin by computing the initial camera rays.
This is done through a standard GPU volume raycasting approach where the backfaces of the volume's bounding box
are rasterized and ray directions computed
in the fragment shader~\cite{stegmaier_simple_2005,engel_real-time_2006}.
The fragment shader writes the pixel's ray direction and the $t$ value
that it enters the volume out to an image-sized ray data buffer,
requiring 16 bytes per-ray.
Rays that miss the volume are marked as terminated.


\subsection{Macrocell Grid Traversal}
\label{sec:method_macrogrid_traversal}
Each pass of the wavefront ray traversal begins by finding the next block along the ray that potentially contains
the isosurface (\Cref{fig:method_pipeline_sketch}a).
We traverse the two-level macrocell grids using the algorithm of Amanatides and Woo~\cite{amanatides_fast_1987},
skipping cells whose value range does not contain the isovalue.
Rays begin by traversing the coarse grid. When a coarse cell containing the isovalue is
encountered, we traverse the $4^3$ grid of its blocks to determine if
the ray intersects a block containing the isovalue.
If such a block is found, we record the block ID for the ray in $R_{BID}$, save the coarse and fine grid iterator
traversal states, and exit the macrocell grid traversal kernel.
$R_{BID}$ is an image-sized buffer that stores the block ID each ray intersects,
or \texttt{UINT\_MAX} if none.
Rays that exit the volume are marked as terminated.
The macrocell grid traversal is run over all $w \times h$ rays; rays that have terminated simply
early exit from the kernel.

The grid iterator states are saved and restored between passes to ensure that
we do not skip cells due to precision issues that would occur when simply tracking the ray's current $t$ value.
Iterator states are stored in an image-sized buffer that tracks $t_{\text{max}}$ and the current cell ID,
requiring 16 bytes per-grid for a total of 32 bytes per-ray.



\subsection{Mark Visible and Active Blocks}
\label{sec:method_mark_blocks}
\begin{figure*}
    \centering
    \begin{subfigure}{0.32\textwidth}
        \centering
        \includegraphics[width=\textwidth]{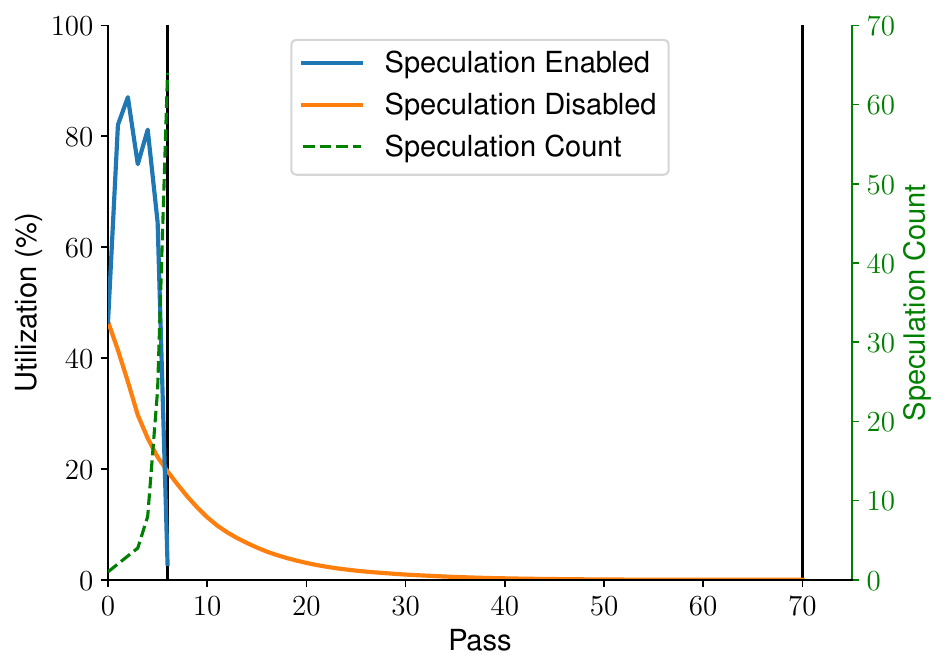}
        \vspace{-1.5em}
        \caption{\label{fig:method_utilization_tacc_plot}%
        Spec. Enabled: 7 passes, 1269ms, \\
        Spec. Disabled: 71 passes, 6025ms.}
    \end{subfigure}
    \begin{subfigure}{0.32\textwidth}
        \centering
        \includegraphics[width=\textwidth]{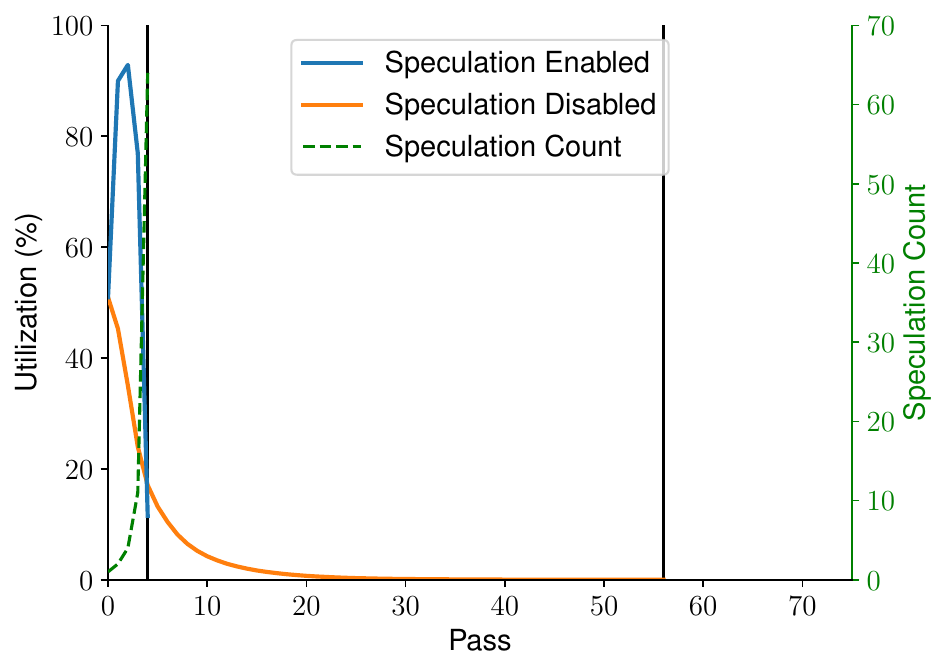}
        \vspace{-1.5em}
        \caption{\label{fig:method_utilization_plasma_plot}%
        Spec. Enabled: 5 passes, 922ms, \\
        Spec. Disabled: 57 passes, 4527ms.}
    \end{subfigure}
    \begin{subfigure}{0.32\textwidth}
        \centering
        \includegraphics[width=\textwidth]{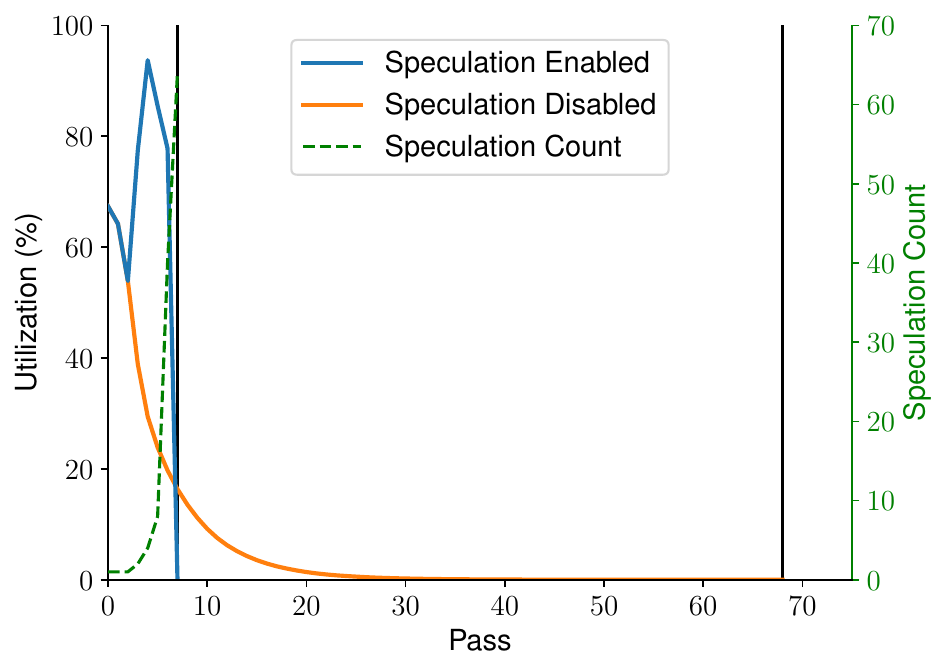}
        \vspace{-1.5em}
        \caption{\label{fig:method_utilization_miranda_plot}%
        Spec. Enabled: 8 passes, 1466ms, \\
        Spec. Disabled: 69 passes, 6892ms.}
    \end{subfigure}
        \vspace{-0.5em}
    \caption{\label{fig:method_utilization}%
    Our speculative ray traversal improves GPU utilization to
    reduce the number of passes needed to render the isosurface by $10\times$ on average,
    thereby reducing the total time to complete the isosurface by $4.8\times$ on average.
    Although average time per pass roughly doubles, this is more than made up for by the
    reduction in the total number of passes required.
    The vertical black lines mark when the surface was completed for each configuration.
    The dotted green line shows the speculation count, which is increased as rays terminate to
    process additional speculated ray-block intersections for the remaining active rays in parallel to terminate them sooner.
    Timings are reported on an RTX~3080.}
        \vspace{-1.5em}
\end{figure*}

Next, we determine which blocks need to be decompressed to
process ray-block intersections (\Cref{fig:method_pipeline_sketch}b).
A block is marked both visible and active if a ray is traversing it (\Cref{fig:method_pipeline_sketch}b, green blocks);
blocks that are $+x/y/z$ neighbors of visible blocks must also be decompressed to provide
data for the visible block's dual grid, and are marked active (\Cref{fig:method_pipeline_sketch}b, blue blocks).
This pass is run on the GPU over the entire $R_{BID}$ buffer, and thus scales with
the image size rather than the number of blocks.
Kernel invocations for terminated rays simply exit early.



\subsection{GPU-driven LRU Block Cache}
\label{sec:method_lru_cache}
The buffer marking active blocks, $M_\text{BAct}$, is passed to the GPU-driven LRU block cache
of Usher and Pascucci~\cite{usher_interactive_2020} to produce a list of the new blocks that need
to be decompressed and cached for the current pass (\Cref{fig:method_pipeline_sketch}c).
These blocks are decompressed into their assigned cache slots using a
WebGPU port~\cite{usher_interactive_2020} of ZFP's~\cite{lindstrom_fixed_rate_2014}
CUDA fixed-rate decompression algorithm.
Rays are likely to require data from
the same blocks traversed in the previous few passes.
The data from these blocks will be readily available in the cache, reducing the decompression
cost for the pass.
Similarly, rays are likely to require data from the same blocks as their neighbors in a given pass.
Shared blocks will be decompressed once and cached, amortizing the decompression
workload over multiple rays.

Performing the cache update each pass allows us to replace unneeded blocks with new ones
each pass, reducing the algorithm's working set to just the active blocks
in an individual pass.
This is in contrast to surface extraction based methods~\cite{usher_interactive_2020},
which decompress and store all the blocks that potentially contain the
isosurface at once, regardless of visibility.


\subsection{Build Raytracing Kernel Inputs}
\label{sec:method_compact_sort_ray_block_ids}
At this point, we have all the volume and ray data
required to traverse rays through the blocks they intersect and test for ray-isosurface intersections.
However, a large number of rays will likely traverse the same block in each pass.
If we were to run the raytracing kernel in parallel over the rays we would
waste bandwidth by repeatedly re-loading the same block from memory.
Instead, we run the raytracing kernel in parallel over the visible blocks.
The raytracing kernel then loads each block's dual grid from memory just once and
computes ray-isosurface intersections for the rays passing through it.

The inputs to the raytracing kernel are the list of visible block IDs ($I_\text{BVis}$),
the number of rays intersecting each block ($N_\text{BRays}$), the offsets to the block's set of rays ($O_\text{BRays}$),
and the active ray IDs sorted by their block ID ($I_\text{RAct}$).
These inputs are produced through a series of stream compactions, prefix sums,
and parallel sorts on the GPU (\Cref{fig:method_pipeline_sketch}d).
The list of visible block IDs, $I_\text{BVis}$, is computed via a stream compaction.
The number of rays intersecting each block, $N_\text{BRays}$, is computed using a kernel run for
each ray that atomically increments the block's ray count.
The offset to each block's set of ray IDs, $O_\text{BRays}$, is computed by perfoming
a prefix sum on $N_\text{BRays}$.
Finally, we compute the list of active ray IDs ($I_\text{RAct}$) sorted by
their block ID ($I_\text{RActBID}$) by compacting the active ray IDs and their block IDs,
then performing a parallel sort by key, using the block ID as the key.

\subsection{Raytracing Visible Blocks}
\label{sec:method_rt_visible_blocks}
The raytracing kernel is run in parallel over the visible blocks, and is responsible
for taking the set of rays intersecting the block and traversing them through its
dual grid to find ray-isosurface intersections (\Cref{fig:method_pipeline_sketch}e).
The kernel consists of two steps: loading the block's dual grid data into shared memory,
followed by traversing the rays through the dual grid to compute intersections.

The block's dual grid consists of its local data combined with the face/edge/corner values
from its neighbors in the $+x/y/z$ direction, if those neighbors exist.
We employ the parallel loading strategy of Usher and Pascucci~\cite{usher_interactive_2020}
to load the dual grid data into shared memory.
Kernel work groups are launched with 64 threads, corresponding to one thread
per dual grid cell, and have a work group shared memory region with room for $5^3$ floating point values
to store the full set of local and neighbor values for the dual grid.
First, the work group loads the 64 vertices corresponding to the block's local $4^3$ data
into the shared memory region,
after which a subset of threads load data from the $+x/y/z$ face, edge, and corner neighbor
blocks to complete the dual grid.
Finally, the work group synchronizes on a memory barrier to ensure the complete
dual grid data is visible to all threads in the group.

With the dual grid loaded into shared memory, we can now traverse rays through it
to find ray-isosurface intersections.
The 64 threads in the work group are used to process the block's rays in parallel in chunks of 64 rays,
with each thread responsible for a different ray in the chunk.
We again use the Amanatides and Woo~\cite{amanatides_fast_1987}
grid traversal algorithm to step rays through the dual grid.
Ray-isosurface intersections are computed using the ray-voxel intersection technique
of Marmitt et al.~\cite{marmitt_fast_2004}.
If an intersection is found, the shaded color and depth is output to the ray's pixel
in the framebuffer and the ray is marked as terminated.



\section{Increasing GPU Utilization with Speculation}
\label{sec:speculation}

Our algorithm as described in~\Cref{sec:method} achieves interactive isosurface rendering of massive
data sets within a small memory footprint. However, we observed that the algorithm would take a large number
of passes to complete the isosurface on average.
Each pass incurs some fixed time costs, and this translated into long total surface rendering times.
We further observed that, on average, after 10 passes there were $<20\%$ of rays still active, and that by pass 25
there were $<1\%$ of rays still active (see~\Cref{fig:method_utilization}).
These long tail rays are those that just miss the surface and must be traversed through many blocks
before finding an intersection or exiting the volume. 

To address this issue, we extend our algorithm to enable \textit{speculative} intersection of rays
with additional blocks to increase utilization and terminate rays in fewer passes (\Cref{fig:method_traversal_speculation}).
To avoid scaling up memory consumption and compute costs by the speculation count,
we treat the various image-sized ray and block data buffers
used by our algorithm as a virtual GPU with $w \times h$ threads and memory slots.
As rays terminate, these slots become available for other active rays to use for speculation.
For simplicity we use a constant speculation count for all rays,
defined as $N_\text{Spec} = \lfloor\frac{w \times h}{N_\text{Act}}\rfloor$,
where $N_\text{Act}$ is the number of active rays.
To balance between terminating rays in fewer passes and performing unnecessary computation,
we limit the speculation count to a maximum of 64.


The following modifications are made to the algorithm described previously (\Cref{sec:method}) to enable speculation.
The macrocell grid traversal kernel now advances each ray through $N_\text{Spec}$ blocks,
recording multiple block IDs for each ray (\Cref{sec:method_speculate_macrocell_traverse}).
As the macrocell grid traversal will write out the same ray ID $N_\text{Spec}$ times in $R_\text{ID}$,
ray IDs in the buffer are no longer unique identifiers, and we must introduce an additional
speculated ray-block offset buffer to the raytracing kernel inputs (\Cref{sec:method_speculate_build_rt_inputs}).
To prevent speculated ray-block intersections from trampling each other's results,
the block raytracing kernel is modified to write intersection results out to a new RGBZ buffer instead of directly
to the framebuffer (\Cref{sec:method_speculate_rt_blocks}).
A new kernel is introduced to select the closest hit found, if any, for a given ray
and write the final color to the framebuffer (\Cref{sec:method_speculate_depth_composite}).
At the end of each pass, we keep the prefix sum result buffer $O_\text{Act}$ 
that is produced when computing $N_\text{Act}$ and update $N_\text{Spec}$.
$O_\text{Act}$ is used to assign offsets in
$R_{ID}$ and $R_{BID}$ to the remaining active rays.




\subsection{Speculative Macrocell Grid Traversal}
\label{sec:method_speculate_macrocell_traverse}
Our speculative macrocell grid traversal performs the same traversal
as before (\Cref{sec:method_macrogrid_traversal}), with the key difference
being that it traverses the ray until finding up to $N_\text{Spec}$ visible blocks
instead of just one (see~\Cref{fig:method_traversal_speculation}), and records
all the visible block IDs encountered to be tested for intersections.
The set of blocks being traversed by a given ray may be disconnected
due to empty space-skipping.

The macrocell grid traversal kernel is run over all $w \times h$ pixels as before,
with terminated rays exiting early.
The $N_\text{Spec}$ entries for each active ray are written at offsets
given by $o = O_\text{Act}[\text{ray}] \times N_\text{Spec}$.
The visible block IDs for each active ray are written into $R_\text{BID}$
starting at $o$, with up to $N_\text{Spec}$ entries written for each ray.
If the ray exits the volume early, its remaining $R_\text{BID}$ entries are left
filled with \texttt{UINT\_MAX} and filtered out in subsequent passes in the manner
as terminated rays.
The ray ID buffer, $R_\text{ID}$, is populated by writing out
$N_\text{Spec}$ entries of the ray ID starting at $o$.
As before, each ray maintains just one coarse and fine grid iterator state.
The iterator states are saved out after $N_\text{Spec}$ visible blocks have been found,
to resume traversal after the last block being intersected in the pass.

As each speculated ray-block intersection writes its block ID to the $R_\text{BID}$
buffer as before, the mark visible and active blocks kernel does not require
modification to support speculation.
The kernel is run over the entire $R_\text{BID}$ buffer and marks blocks active as before,
with the only difference being that some visible block IDs in the buffer correspond to
speculated ray-block intersections.


\subsection{Build Speculated Raytracing Kernel Inputs}
\label{sec:method_speculate_build_rt_inputs}

The construction of the inputs for the raytracing kernel when speculation is enabled
is nearly identical to the step without speculation (\Cref{sec:method_compact_sort_ray_block_ids}).
The key difference is that ray IDs are now repeated $N_\text{Spec}$ times
in the active ray ID buffer $I_\text{RAct}$, meaning that
the ray ID alone is no longer a unique identifier for a ray-block intersection.

We introduce an additional offset buffer, $O_\text{Spec}$,
that assigns a unique index to each ray-block intersection.
$O_\text{Spec}$ is produced by scanning the buffer that marks active ray-block intersections, $M_\text{RAct}$.
$M_\text{Ract}$ is produced as before during the compaction of active ray IDs (\Cref{fig:method_pipeline_sketch}d.3).
As with $I_\text{RAct}$, $O_\text{Spec}$ is compacted down to just the entries
for active ray-block intersections and sorted by block ID to match the order of $I_\text{RAct}$.
The list of visible block IDs ($I_\text{BVis}$), the number of rays to
process for each block ($N_\text{BRays}$), and the offsets ($O_\text{BRays}$)
are produced as before.



\subsection{Raytracing Visible Blocks with Speculation}
\label{sec:method_speculate_rt_blocks}
With the entries in $I_\text{RActive}$, $I_\text{BVis}$, $N_\text{BRays}$ and $O_\text{BRays}$
already accounting for speculated ray-block intersections, few modifications
are needed to the raytracing kernel.
As before, after loading the dual grid data each visible block reads its
ray IDs from the offset given in $O_\text{BRays}$ and traverses the rays
through its dual grid to find ray-isosurface intersections.
However, as intersections may be found in multiple blocks for a given ray when speculation
is enabled, the kernel is modified to output intersection results to a new RGBZ buffer instead of
directly to the framebuffer.
Color and depth values for ray-isosurface intersections are written at offsets
given in $O_\text{Spec}$ for the ray-block intersection.

\subsection{Depth Compositing Speculated Intersections}
\label{sec:method_speculate_depth_composite}

The final step in our speculative rendering pipeline is to perform depth compositing
on the set of intersections found for each ray.
A kernel is run for each active ray that iterates through its $N_\text{Spec}$ potential
intersections to select the closest one, if any, and writes it to the framebuffer.
We note that it would be possible to skip the depth compositing step if WebGPU
supported 64-bit atomics, as the depth sorting could be performed using atomic min
operations in the raytracing kernel instead~\cite{schutz_software_2022}.
Rays that exit the volume without finding a hit are also marked as terminated
in this step.

\section{Evaluation}
\label{sec:evaluation}

\begin{table}
    \centering
    \relsize{-1}{
    \begin{tabular}{@{}cccc@{}}
    \includegraphics[width=0.20\columnwidth]{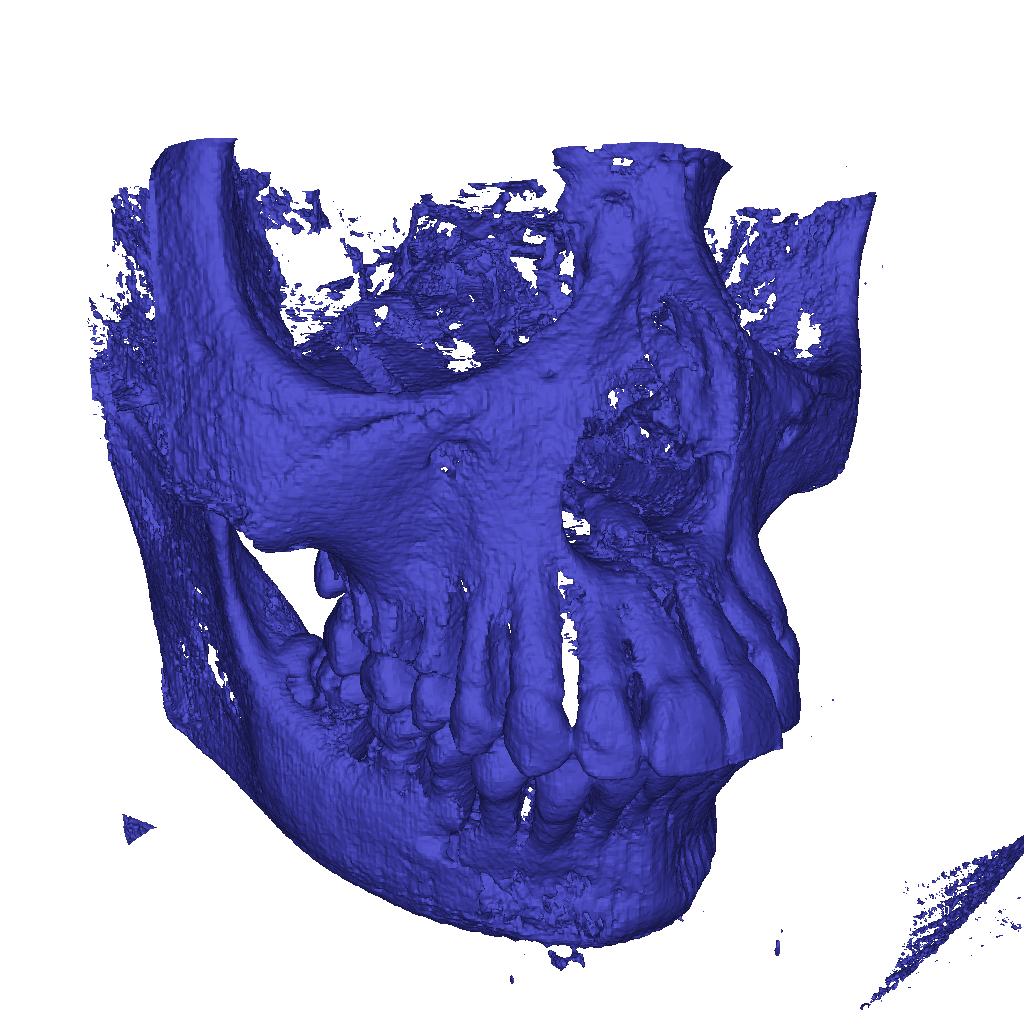} &
    \includegraphics[width=0.20\columnwidth]{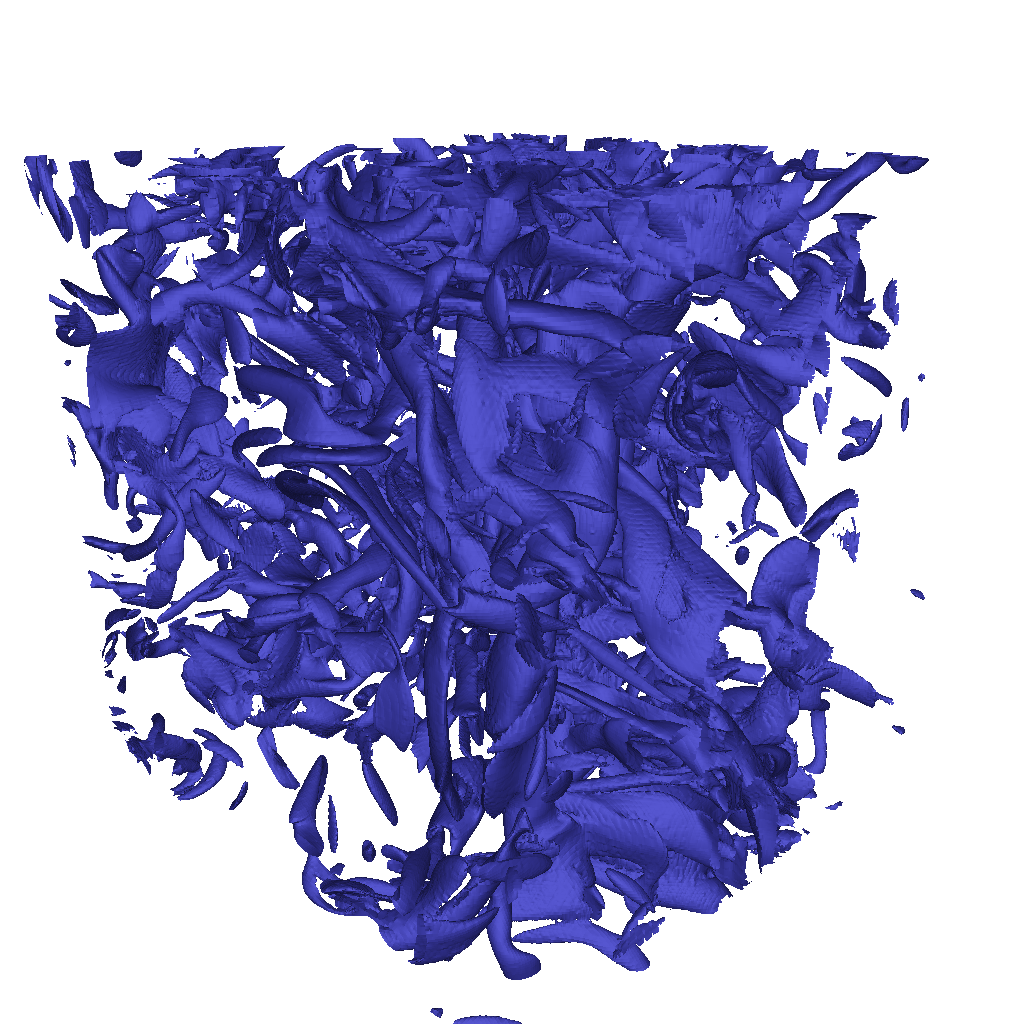} &
    \includegraphics[width=0.20\columnwidth]{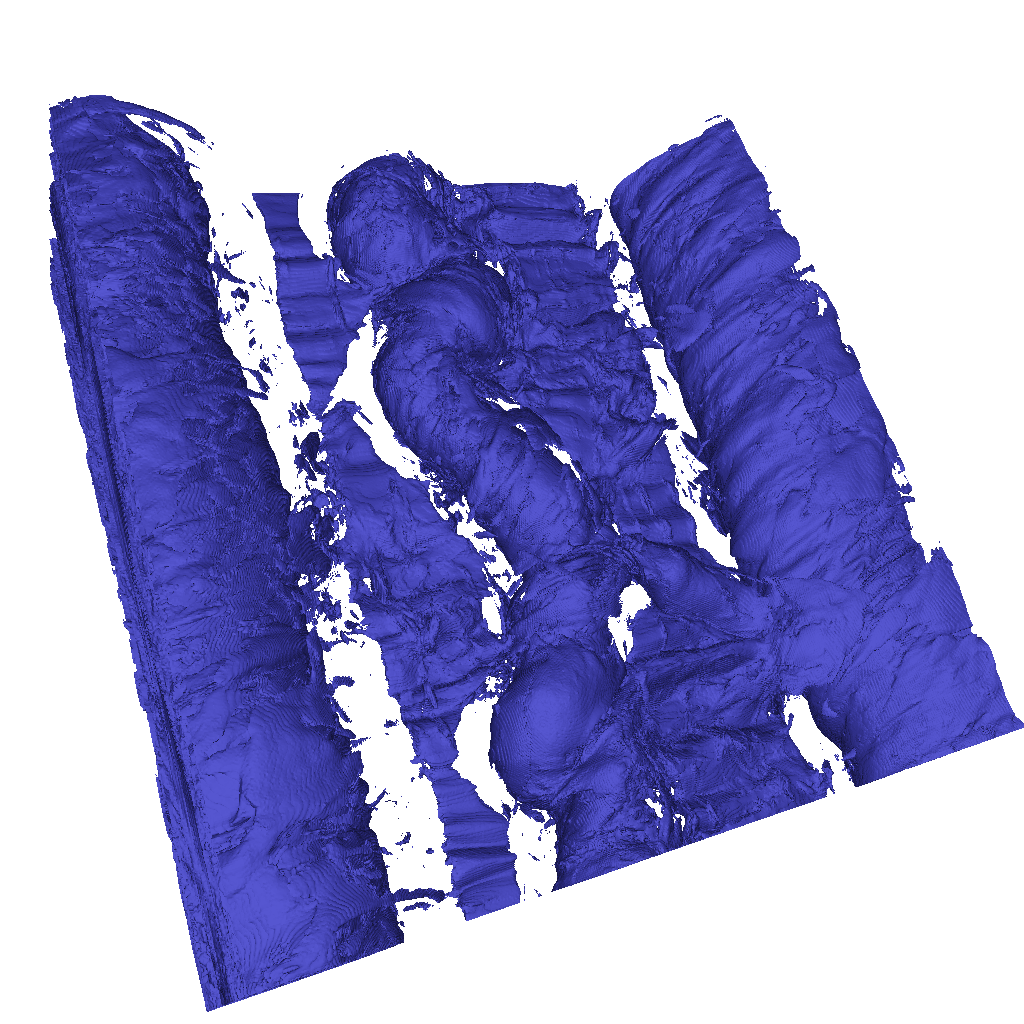} &
    \includegraphics[width=0.20\columnwidth]{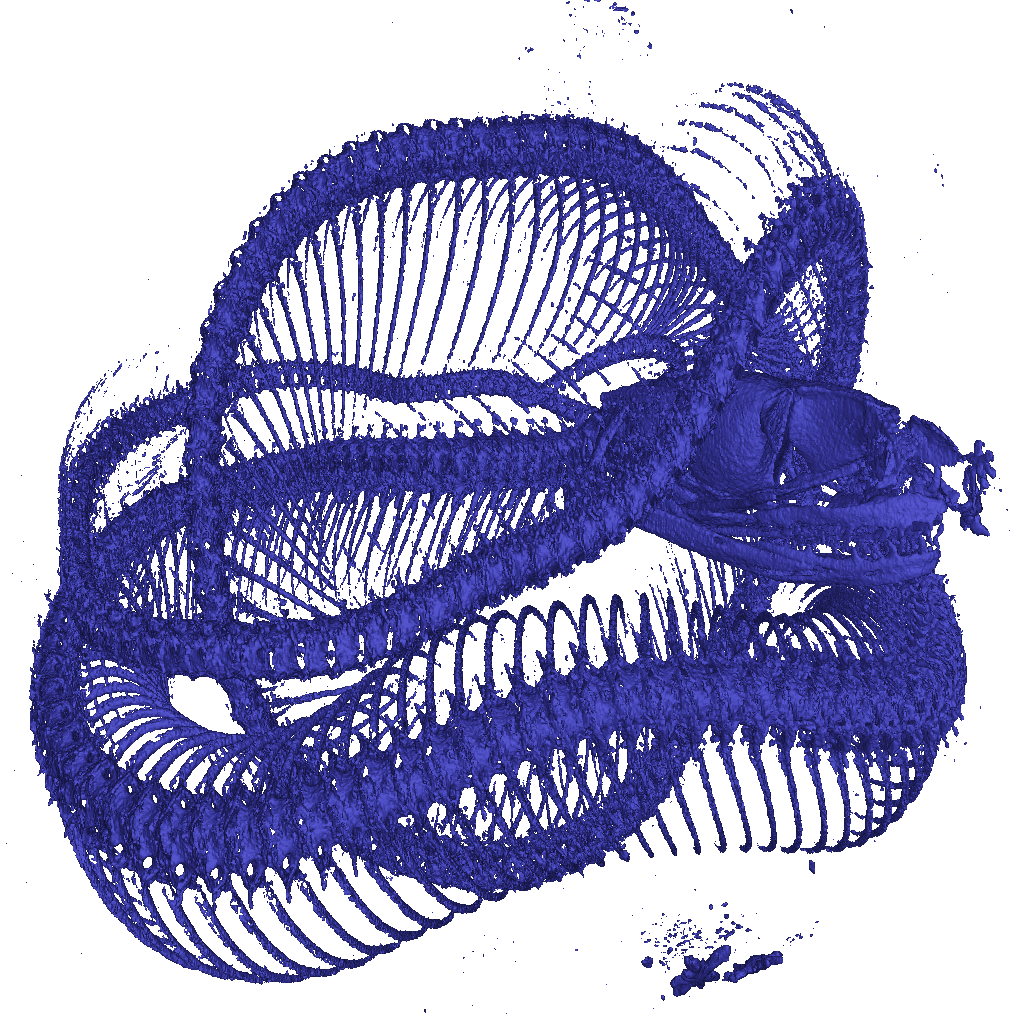} \\
    \textsf{Skull} & \textsf{TACC} & \textsf{Plasma} & \textsf{Kingsnake} \\
    $256^3$ & $256^3$ & $512^3$ & $1024 \times 1024 \times 785$ \\
    \end{tabular}
    \begin{tabular}{@{}ccc@{}}
    

    \includegraphics[width=0.3\columnwidth]{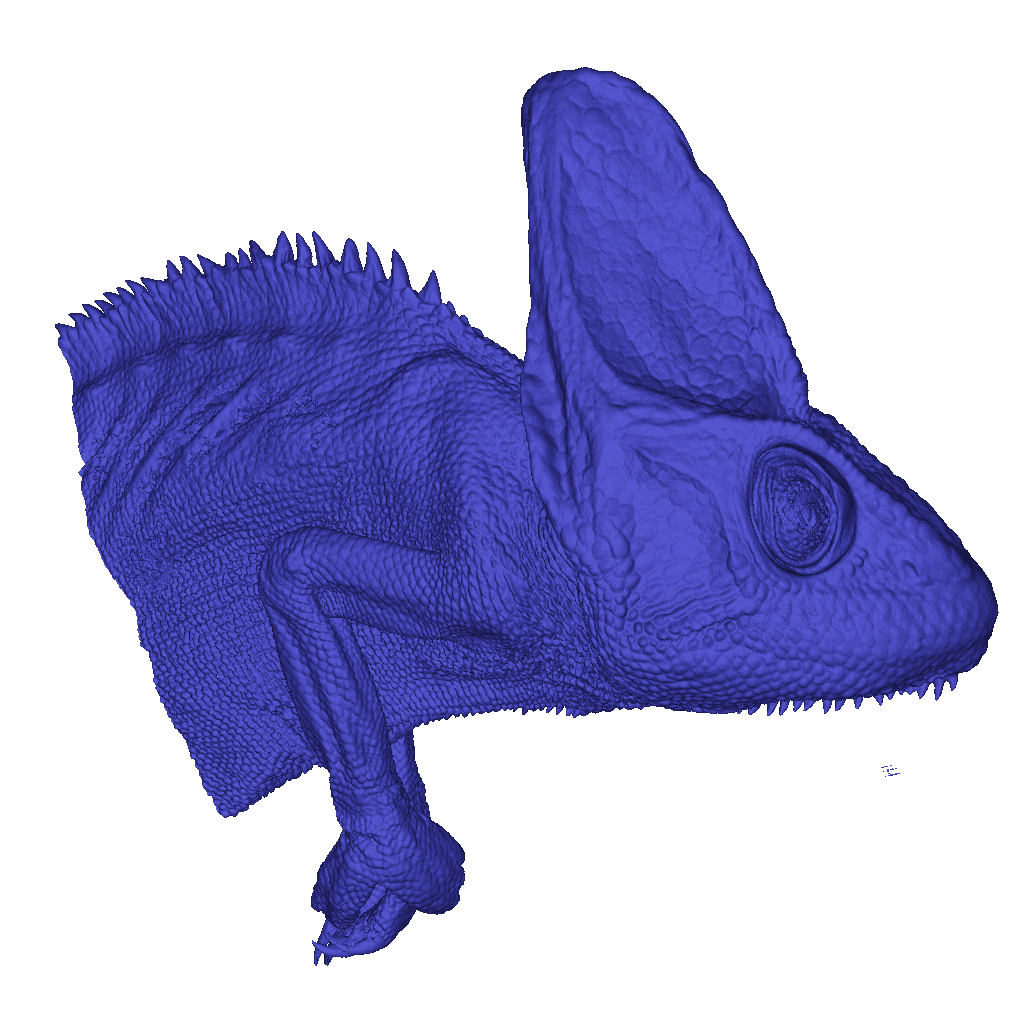} &
    \includegraphics[width=0.3\columnwidth]{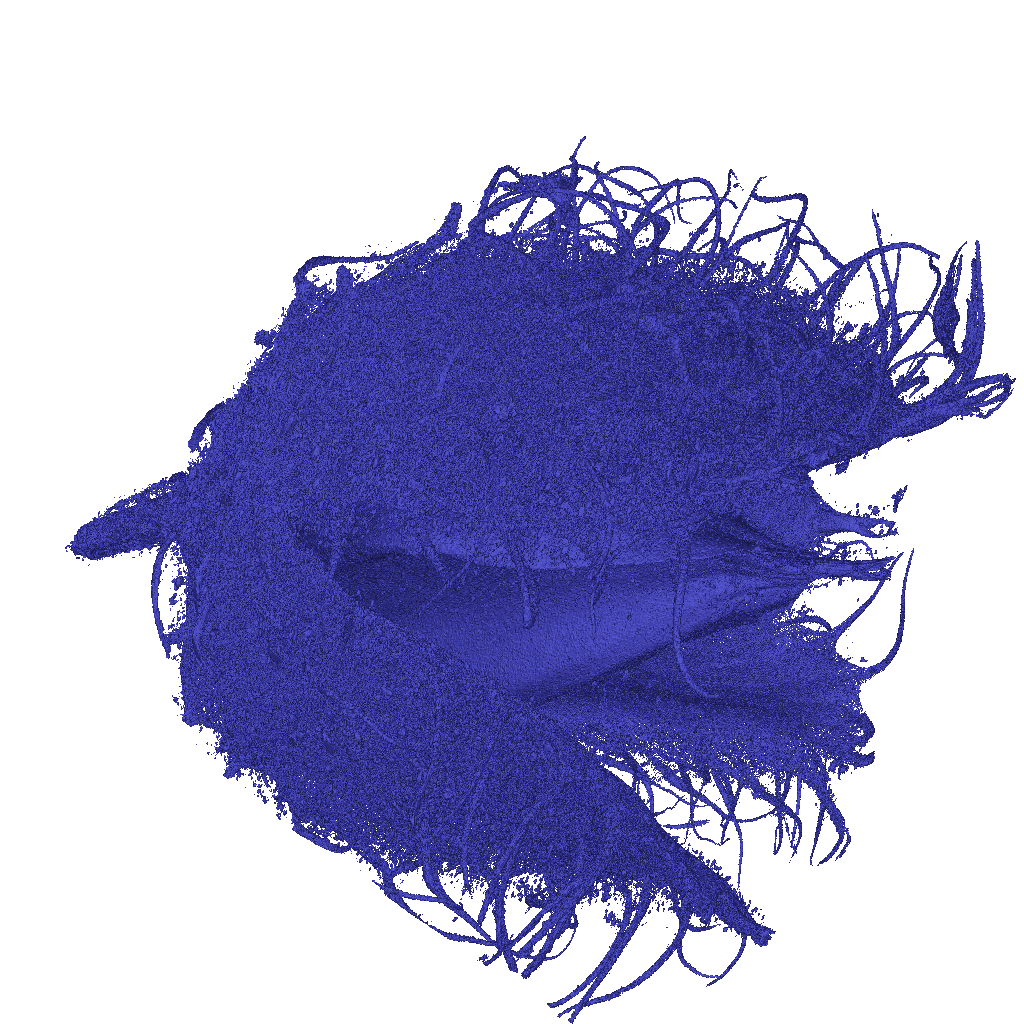} &
    \includegraphics[width=0.3\columnwidth]{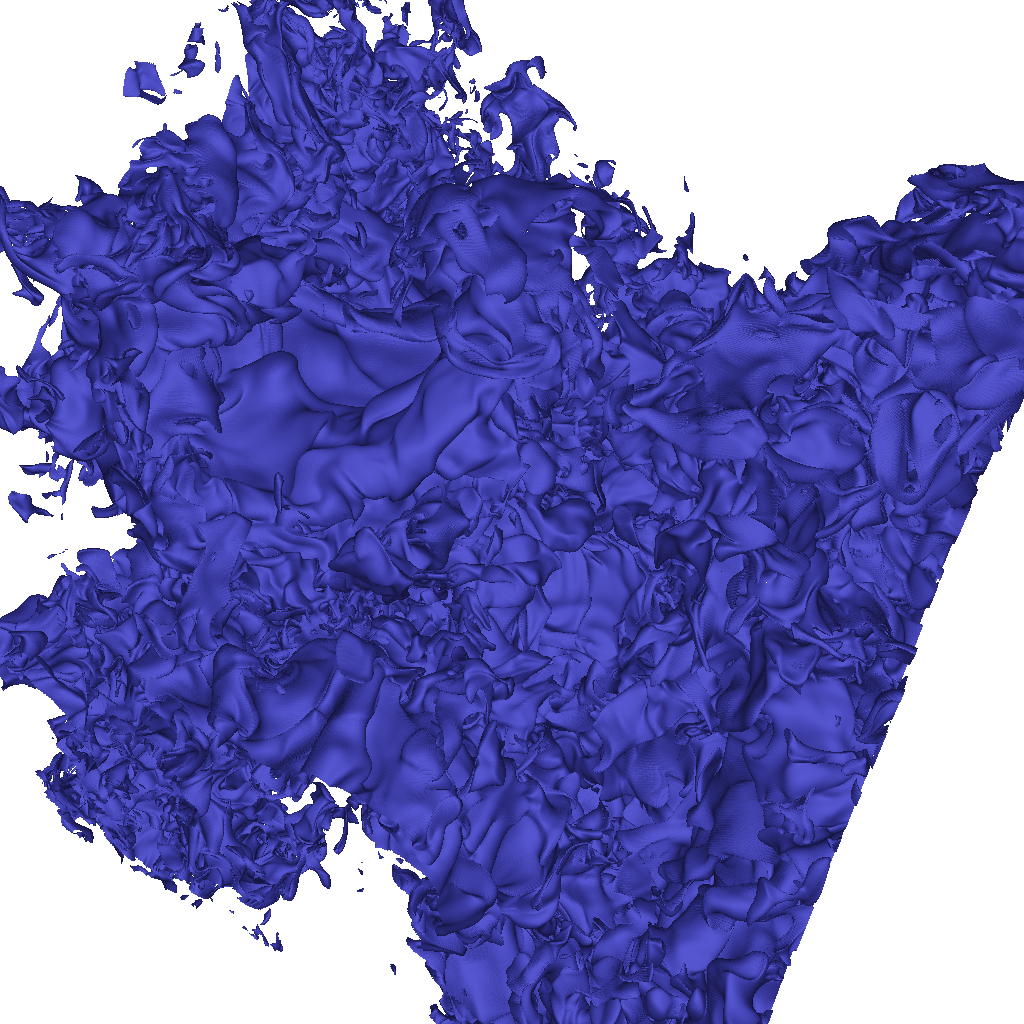} \\
    \textsf{Chameleon} & \textsf{Beechnut} & \textsf{Miranda} \\
    $1024 \times 1024 \times 1080$ & $1024 \times 1024 \times 1546$ & $1024^3$ \\
    
    \includegraphics[width=0.3\columnwidth]{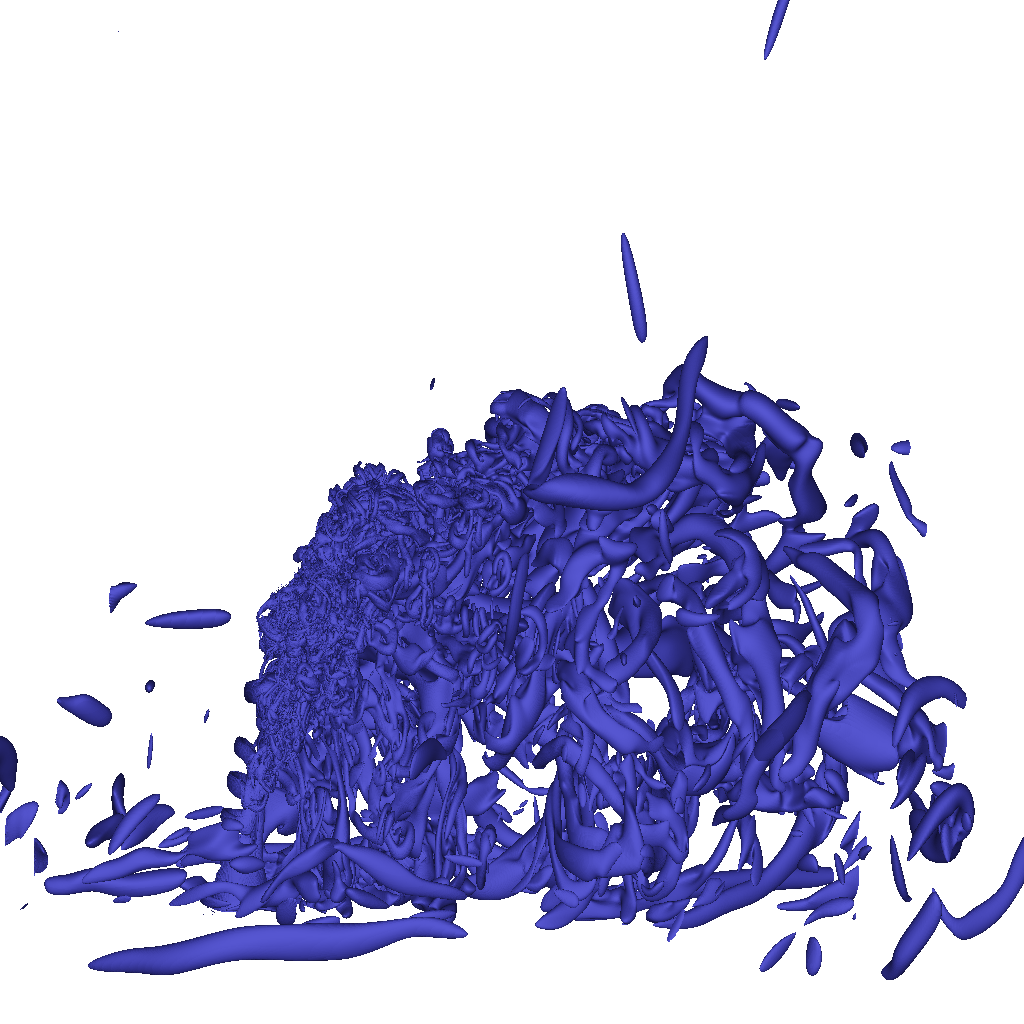} & 
    \includegraphics[width=0.3\columnwidth]{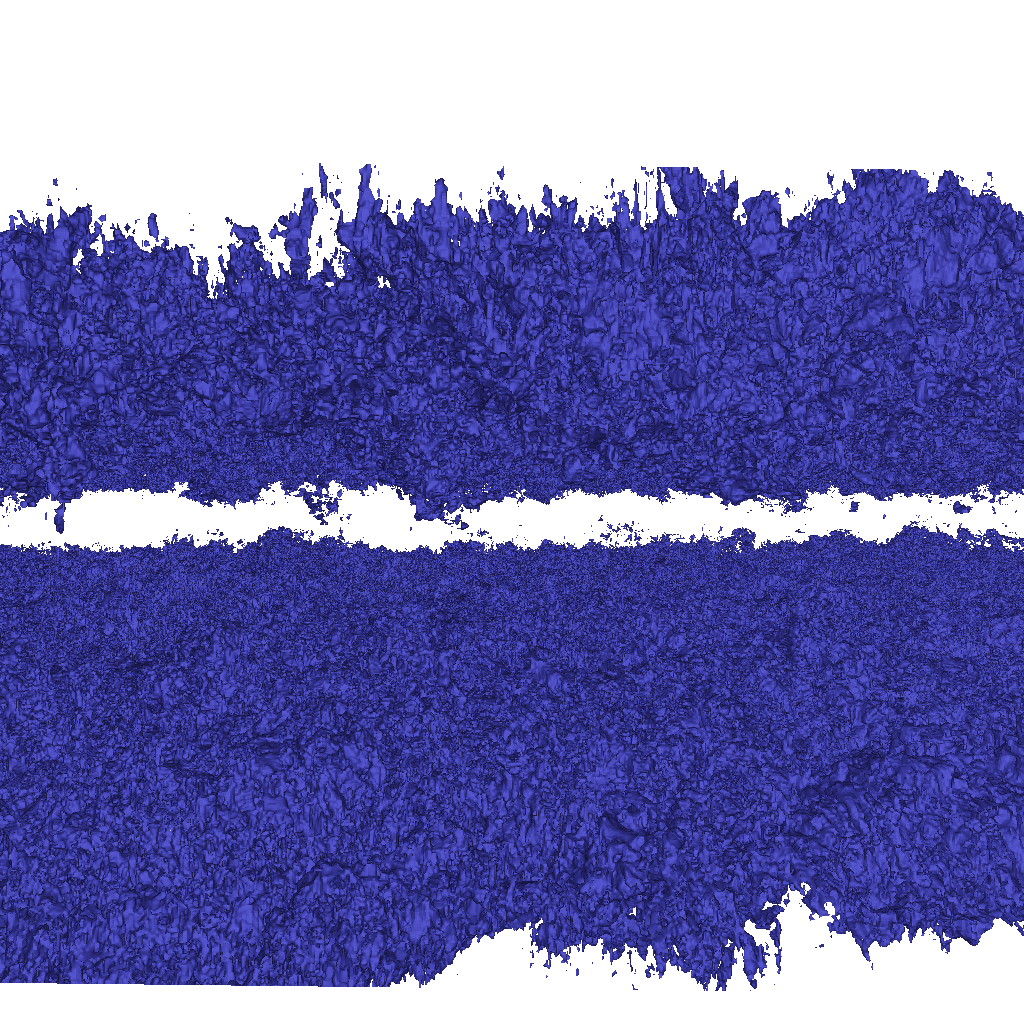} &
    \includegraphics[width=0.3\columnwidth]{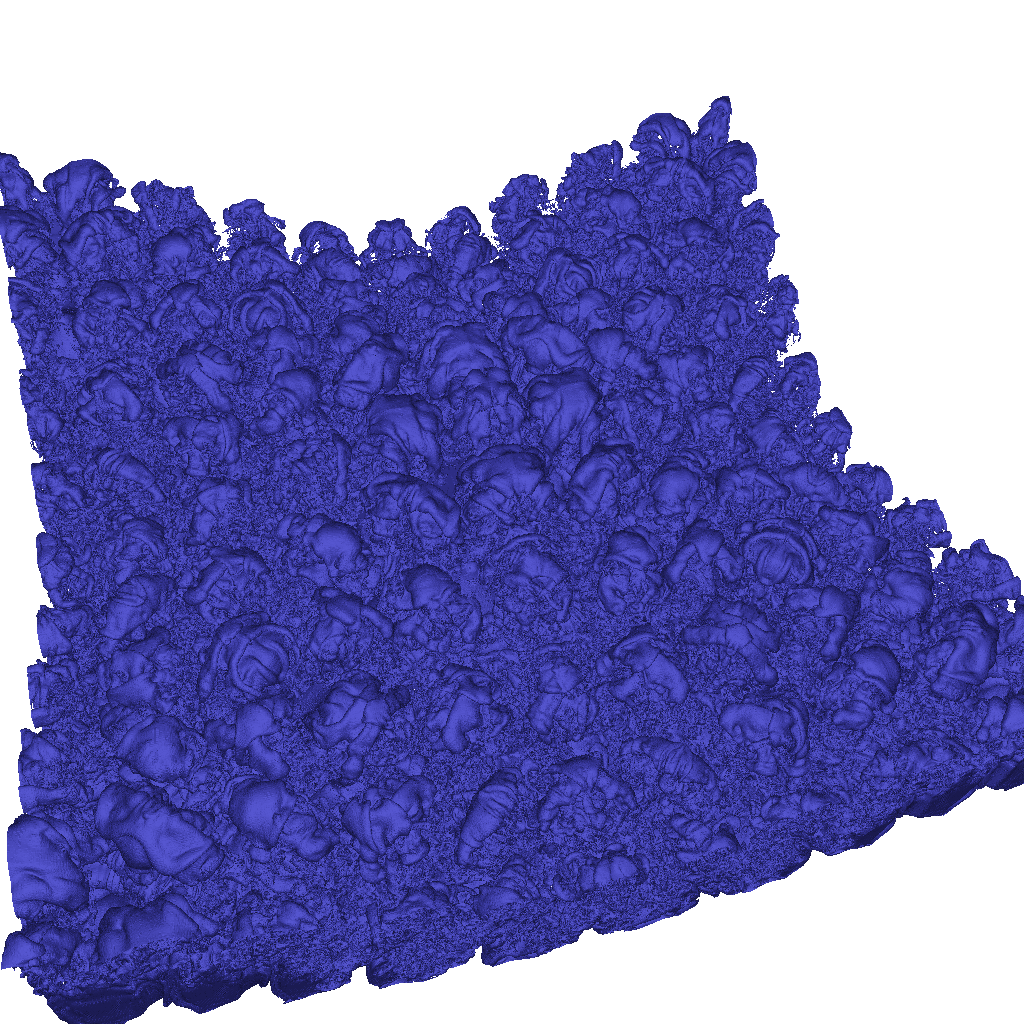} \\
    \textsf{JICF Q} & \textsf{DNS} & \textsf{R-M} \\
    $1408 \times 1080 \times 1100$ & $1920 \times 1440 \times 288$ & $2048 \times 2048 \times 1920$
    \end{tabular}
    }
    \vspace{-1em}
    \caption{\label{tab:all_datasets}%
    The data sets used for evaluation range from small to massive and come from both measured
    data sets and simulations, covering a wide range of isosurface visualization use cases.
    The DNS combines adaptive precision and resolution techniques~\cite{hoang_efficient_2021} to
    enable visualization of the original 1TB ($10240 \times 7680 \times 1536$) volume in the browser.}
    \vspace{-1.5em}
\end{table}

We evaluate the rendering performance and memory consumption of our
method on data sets ranging in size from $256^3$ (16.7M voxels) up
to $2048 \times 2048 \times 1920$ (8.05B voxels) (\Cref{tab:all_datasets}).
Each data set is compressed offline with ZFP to produce the compressed data
used by the renderer. As ZFP only supports single- and double-precision floating
point values, the compression step also converts any non single-precision
data sets to single-precision.
Each data set is benchmarked on 100 random isovalues sampled over a
range covering values of interest in the data.
Each isovalue is rendered over a 10 position camera orbit to a $1280\times720$ framebuffer.
We also demonstrate visualization of complex isosurfaces on the 1TB DNS data set;
the DNS is first resampled from $10240 \times 7680 \times 1536$ to $1920 \times 1440 \times 288$
through a combination of adaptive precision and resolution techniques~\cite{hoang_efficient_2021},
then compressed with ZFP.

The test data sets cover a range of isosurface visualization
scenarios, with some being especially challenging for surface extraction techniques.
The Skull, Kingsnake, Chameleon, and Beechnut were produced through various scanning
technologies.
The Skull and Chameleon consist of relatively smooth shell-like isosurfaces,
while the Kingsnake contains many fine features.
The Beechnut is a challenging case with many fine features and noise, resulting in large isosurface mesh where
large numbers of triangles will be occluded.
The TACC, Plasma, Miranda, JICF~Q, DNS, and Richtmyer-Meshkov (R-M) were produced through various simulation codes.
The Miranda, DNS, and R-M pose similar challenges to surface extraction techniques as the Beechnut;
they consist of highly turbulent isosurfaces that result in large meshes with large numbers of occluded triangles.
The JICF~Q is similarly challenging, as a few isosurfaces cover a substantial portion of the
domain, producing a large surface that requires a large amount of data to be decompressed.

We report performance results of our algorithm on three different systems.
Two are representative of lightweight end user systems: a laptop with
an i7-1165G7 CPU and integrated GPU (XPS~13), and a Mac Mini with an M1 chip (M1 Mac Mini).
The final system is a desktop with an RTX~3080 GPU and an i9-12900k CPU.

\begin{figure}[t]
    \centering
    \begin{subfigure}{\columnwidth}
        \centering
        \includegraphics[width=\textwidth]{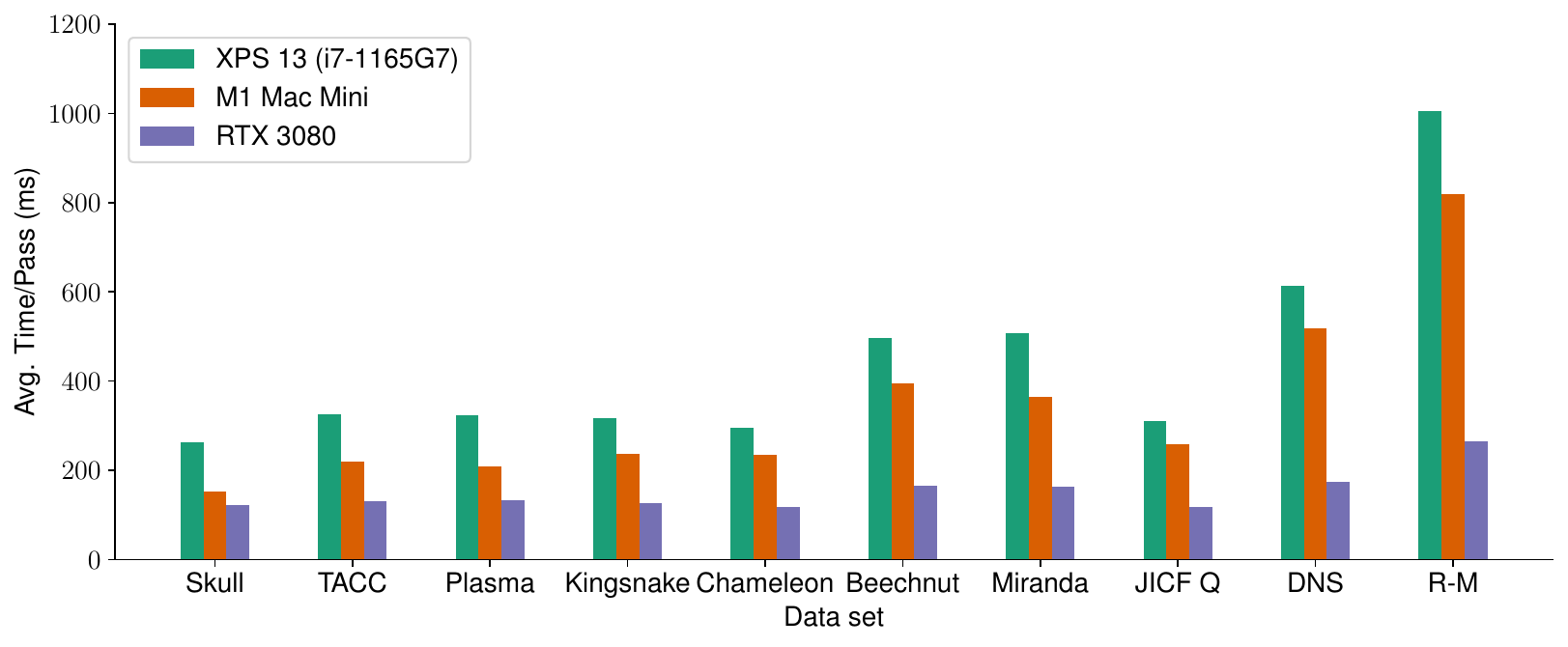}
        \vspace{-1.5em}
        \caption{\label{fig:perf_ours_pass}%
        Average time per pass.}
    \end{subfigure}
    \begin{subfigure}{\columnwidth}
        \centering
        \includegraphics[width=\textwidth]{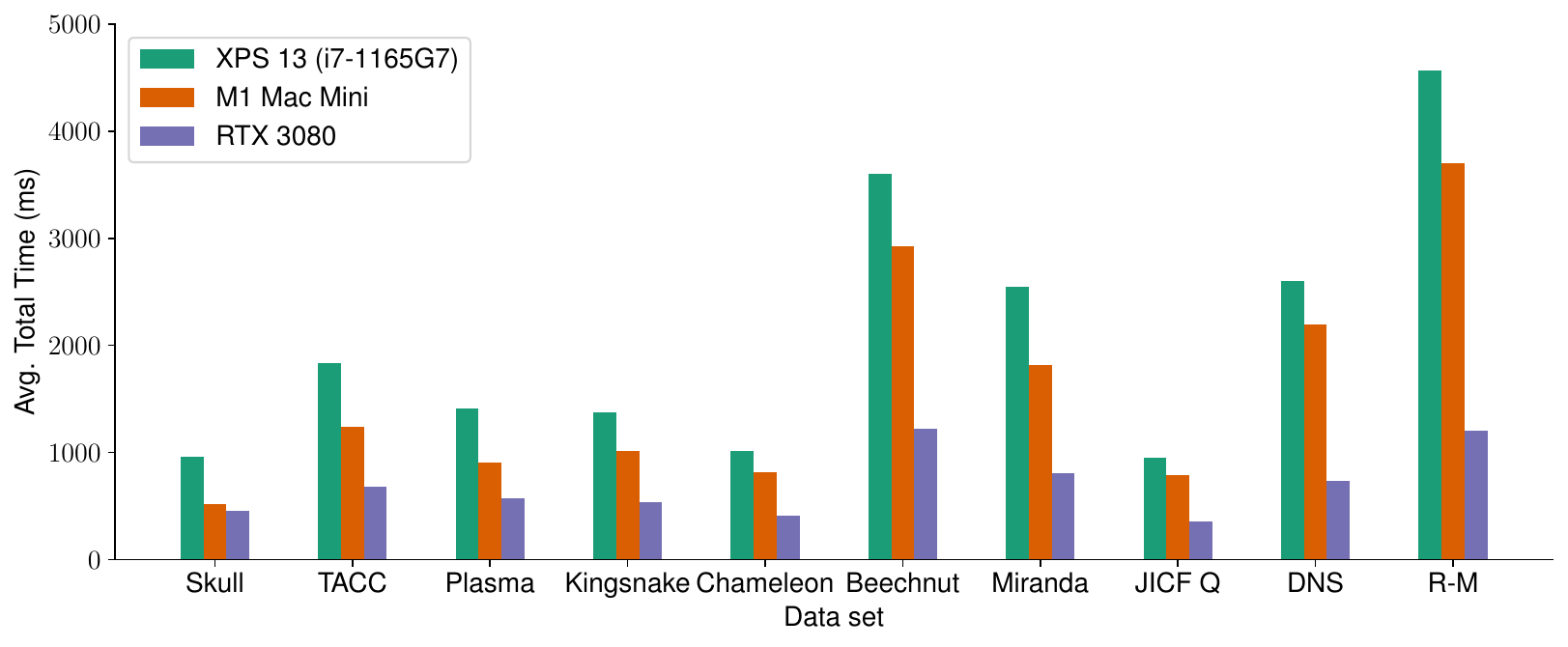}
        \vspace{-1.5em}
        \caption{\label{fig:perf_ours_all}%
        Average total time to complete the isosurface.}
    \end{subfigure}
        \vspace{-1.5em}
    \caption{\label{fig:eval_perf_plots}%
    Our method achieves interactive rendering framerates (a)
    across the data sets tested, even on the XPS~13 and M1 Mac Mini.
    Moreover, rendering cost does not scale significantly with data size, allowing large and complex
    to be rendered interactively on lightweight systems.
    Our speculative approach completes rendering in few passes,
    allowing for reasonable surface completion times (b).}
        \vspace{-1em}
\end{figure}

\begin{figure}
    \centering
    \includegraphics[width=0.95\columnwidth]{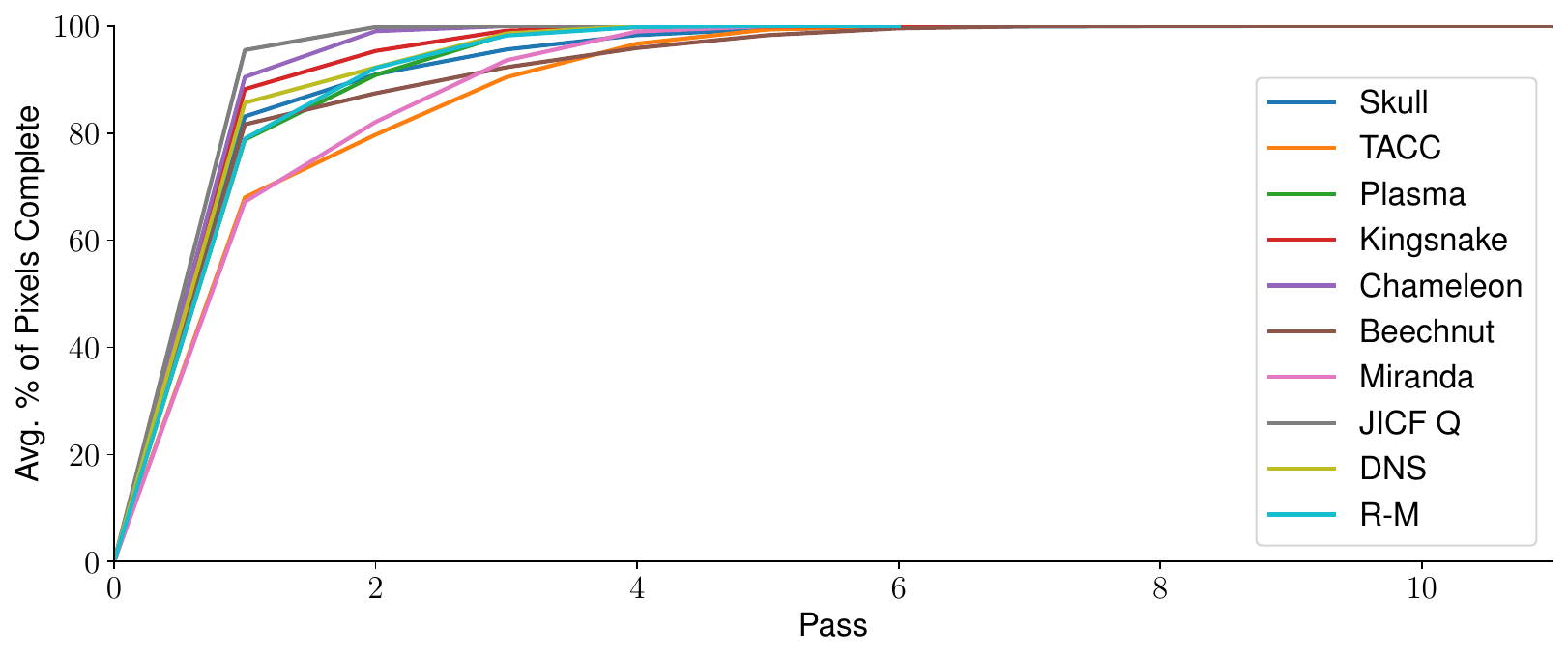}
    \vspace{-1em}
    \caption{\label{fig:eval_image_completeness}%
    Images rendered with our algorithm are over 75\% complete by the second pass.}
    \vspace{-1em}
\end{figure}

We conduct a detailed evaluation of our method's performance and scalability;
and evaluate it against the state of the art in GPU-based large-scale
isosurface extraction~\cite{usher_interactive_2020}.
In \Cref{sec:eval_speculation}, we discuss overall rendering performance
of our method and the benefits of our speculative execution strategy;
in \Cref{sec:eval_scalability}, we evaluate the scalability of
our method with respect to data set size and image resolution compared
to the state of the art.
Finally, \Cref{sec:eval_vs_bcmc} evaluates the memory consumption of our
method against the state of the art.

\subsection{Rendering Performance}
\label{sec:eval_speculation}

The average time per pass and total time to complete the isosurface across
the data sets and hardware platforms tested are shown in~\Cref{fig:eval_perf_plots}.
Our method achieves interactive pass computation times,
and thus rendering frame rates, even when visualizing the massive and complex isosurfaces
of the Beechnut, Miranda, JICF~Q and DNS data sets on the XPS~13 and
M1 Mac Mini.

\Cref{tab:eval_avg_utilization_w_spec} lists statistics about the computation recorded
over the isosurface benchmarks.
Utilization is reported as the percentage of $w \times h$ slots of the virtual GPU
being used for ray-block intersections.
We find that our speculative execution strategy increases GPU utilization to complete
surfaces in far fewer passes, reducing the total time taken to complete
the isosurface (\Cref{fig:method_utilization}, \Cref{fig:perf_ours_all}),
at the cost of slightly increased per-pass times.
The view-dependent nature of our approach allows rendering surfaces with a small memory
footprint, with just $1.3\%$ of blocks visible per-pass on average.

\begin{table}
    \centering
    \relsize{-1}{
    \begin{tabular}{@{}lrrrr@{}}
        \toprule
        \makecell[l]{Data set} & \makecell[r]{Median \\ \# Passes} & \makecell[r]{Avg. Blocks \\ Visible/Pass} & \makecell[r]{Median Spec. \\ Count} & \makecell[r]{Avg. \\ Utilization} \\
        \midrule
        Skull & 3 & 2.09\% & 7 & 30.71\% \\
        TACC & 6 & 4.12\% & 5 & 51.32\% \\
        Plasma & 4 & 1.56\% & 8 & 49.00\% \\
        Kingsnake & 4 & 0.54\% & 13 & 35.68\% \\
        Chameleon & 3 & 0.34\% & 16 & 29.55\% \\
        Beechnut & 7 & 0.64\% & 13 & 54.45\% \\
        Miranda & 5 & 0.83\% & 5 & 60.22\% \\
        JICF Q & 3 & 0.25\% & 23 & 27.15\% \\
        DNS & 4 & 2.61\% & 10 & 53.94\% \\
        R-M & 4 & 0.27\% & 9 & 55.36\% \\
        \bottomrule
    \end{tabular}
    }
    \vspace{-0.5em}
    \caption{\label{tab:eval_avg_utilization_w_spec}%
    Statistics about the number of passes, percentage of blocks visible, speculation
    count, and utilization over the benchmarks.
    Our view-dependent approach reduces memory consumption by requiring just a
    small fraction of the data each pass.}
    \vspace{-1.5em}
\end{table}

When comparing performance across the data sets tested (\Cref{fig:eval_perf_plots}),
we observe that our algorithm's performance is nearly independent of the data set size.
Instead, our scales with the visible surface area and complexity of the isosurface.
We achieve similar performance on data sets with similar
isosurface structure, such as the Skull, Plasma, Kingsnake, Chameleon and JICF Q,
even though these data sets range in size from $256^3$ to $1408 \times 1080 \times 1100$.
These data sets have relatively smooth isosurfaces, where rays can quickly skip empty space to reach
the isosurface and find an intersection.
In contrast, data sets with noisier or more complex isosurfaces such as the Beechnut, Miranda, DNS, and R-M,
see higher rendering times, as more data must be processed for each ray to find an intersection.
We find that our progressive approach is valuable to quickly provide a nearly complete
image of the data set, with 75\% of pixels complete on average by pass two
across the data sets tested (\Cref{fig:eval_image_completeness}).



\subsection{Scalability with Image and Data Size}
\label{sec:eval_scalability}

The performance of our method is primarily driven by the visible surface area and
complexity of the isosurface being rendered, and is less tied to the data set size.
Another main driver of rendering cost in our method is the number of pixels,
allowing rendering performance to be increased by reducing the image size.
This is in line with prior implicit isosurface and volume raycasting techniques,
which have image-order scaling.
Explicit isosurface extraction techniques, such as BCMC~\cite{usher_interactive_2020},
typically extract the complete triangle mesh for the isosurface, including triangles
that will be occluded in the final rendering.
Although extraction techniques are output-sensitive,
they scale with the size of the output isosurface and are more
effected by data set size.

\begin{figure}[t]
    \centering
    \begin{subfigure}{0.5\columnwidth}
        \centering
        \includegraphics[width=\textwidth]{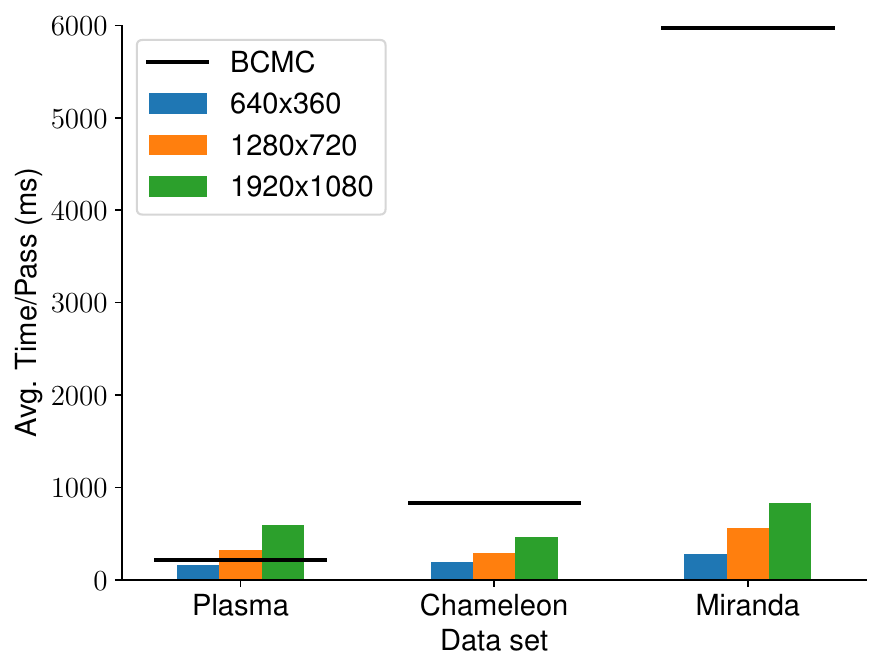}
        \vspace{-1.5em}
        \caption{\label{fig:eval_perf_scaling_xps13_pass}%
        XPS~13, time/pass.}
    \end{subfigure}%
    \begin{subfigure}{0.5\columnwidth}
        \centering
        \includegraphics[width=\textwidth]{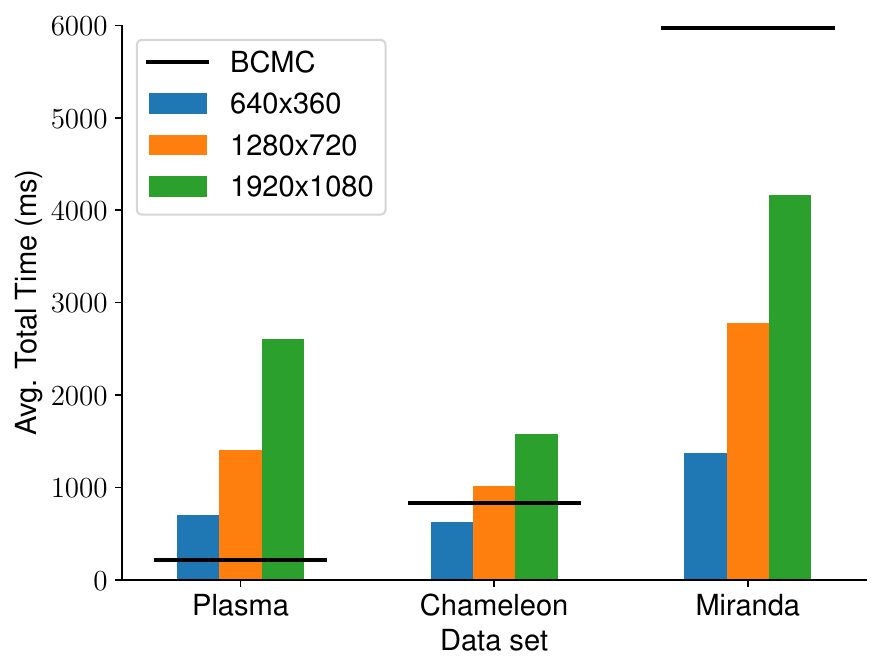}
        \vspace{-1.5em}
        \caption{\label{fig:eval_perf_scaling_xps13_total}%
        XPS~13, total time.}
    \end{subfigure}
    \begin{subfigure}{0.5\columnwidth}
        \centering
        \includegraphics[width=\textwidth]{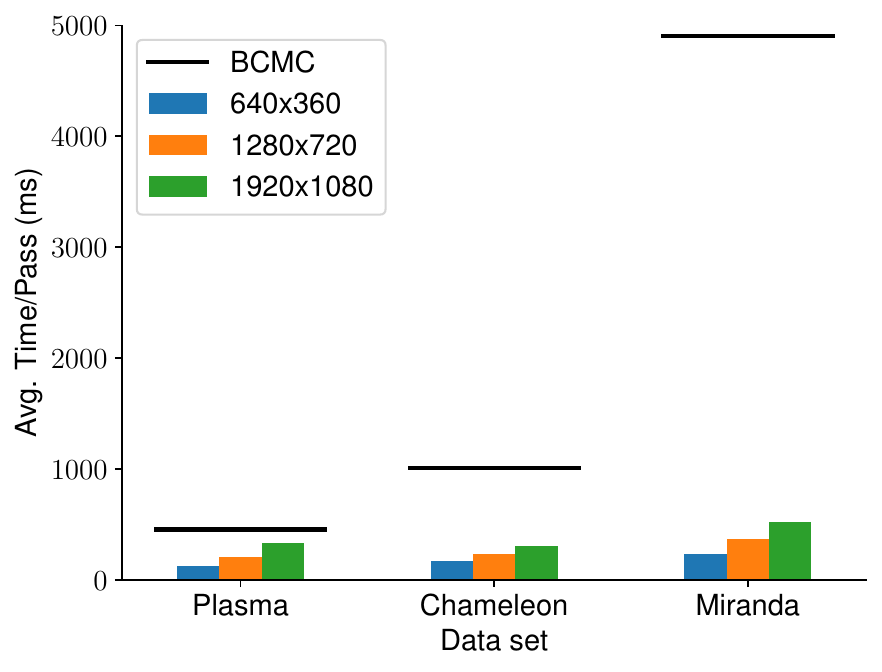}
        \vspace{-1.5em}
        \caption{\label{fig:eval_perf_scaling_m1mini_pass}%
        M1 Mac Mini, time/pass.}
    \end{subfigure}%
    \begin{subfigure}{0.5\columnwidth}
        \centering
        \includegraphics[width=\textwidth]{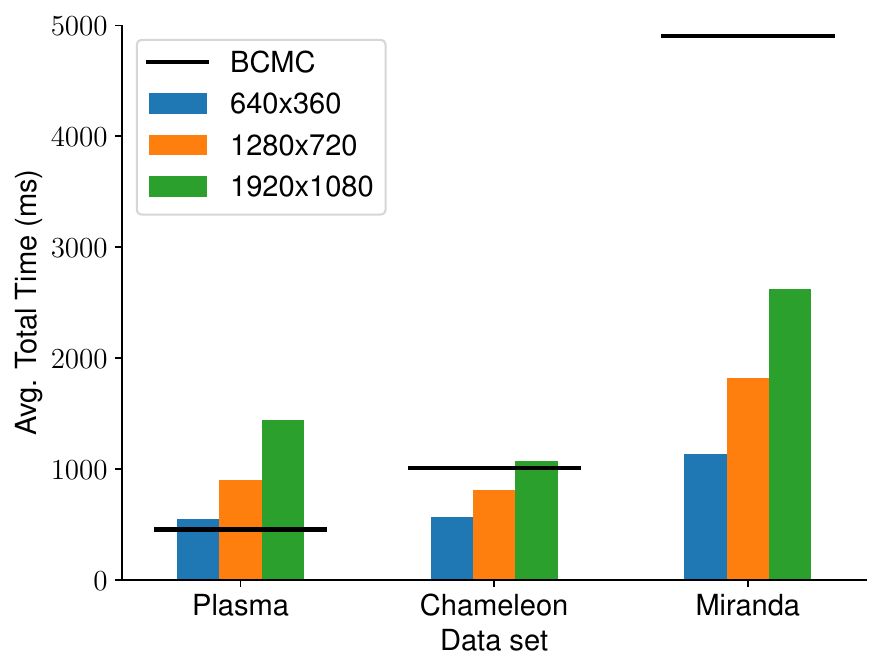}
        \vspace{-1.5em}
        \caption{\label{fig:eval_perf_scaling_m1mini_total}%
        M1 Mac Mini, total time.}
    \end{subfigure}
    \begin{subfigure}{0.5\columnwidth}
        \centering
        \includegraphics[width=\textwidth]{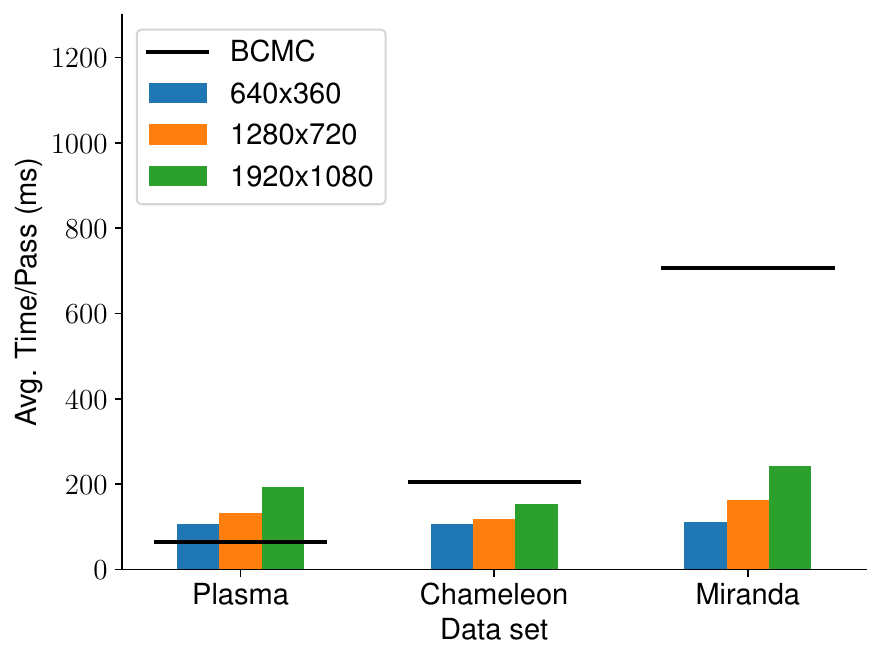}
        \vspace{-1.5em}
        \caption{\label{fig:eval_perf_scaling_3080_pass}%
        RTX~3080, time/pass.}
    \end{subfigure}%
    \begin{subfigure}{0.5\columnwidth}
        \centering
        \includegraphics[width=\textwidth]{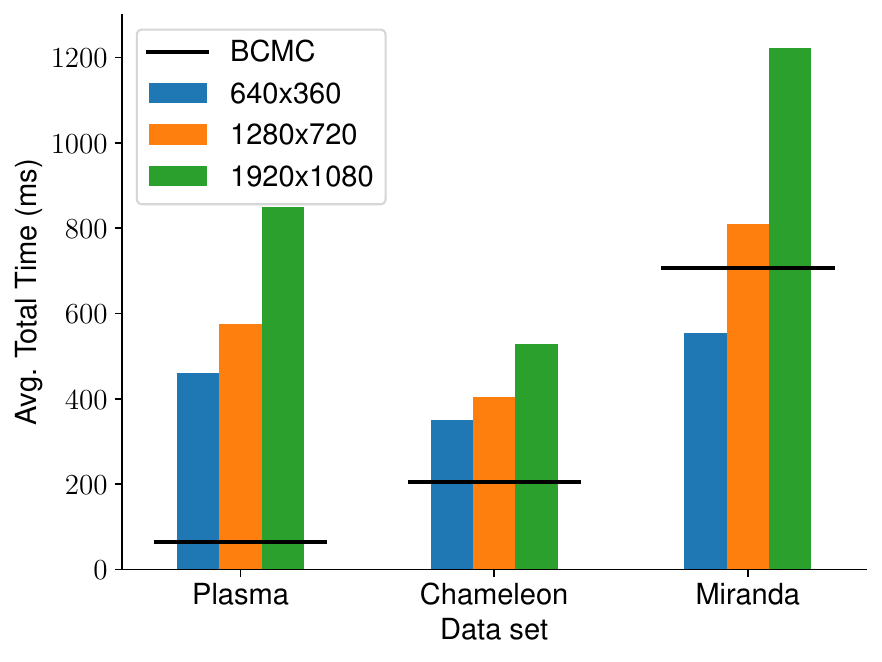}
        \vspace{-1.5em}
        \caption{\label{fig:eval_perf_scaling_3080_total}%
        RTX~3080, total time.}
    \end{subfigure}
        \vspace{-1.5em}
    \caption{\label{fig:eval_perf_scaling}%
    The performance scaling of our approach and BCMC with image resolution and data set size.
    BCMC's compute cost is strongly tied to the size of the data set and the size of the
    isosurface triangle mesh, making it difficult to scale down to ensure interactivity.
    In contrast, our approach can be easily scaled down by reducing
    the image resolution, and is less effected by data size overall, enabling
    interactive rendering of massive data sets on lightweight devices.}
        \vspace{-1.5em}
\end{figure}

\Cref{fig:eval_perf_scaling} quantifies the benefits of these properties of our algorithm
against BCMC~\cite{usher_interactive_2020}.
We conduct benchmarks rendering at $1920 \times 1080$ (1080p), $1280 \times 720$ (720p),
and $640 \times 360$ (360p) on the Plasma, Chameleon, and Miranda
data sets, and compare the average pass and total times of our method
against the isosurface extraction times achieved by BCMC.
Benchmarks for both methods were run over 100 random isovalues.
As before, rendering performance of our method is measured over a 10 position camera orbit
for each isovalue.
Results for our method are shown for each resolution, BCMC is shown as a solid line
as its compute costs are resolution independent.
Our method achieves a $1.50\times$ reduction in per pass and total
times when scaling down from 1080p to 720p,
and an additional $1.55\times$ reduction when scaling down from 720p to 360p.



We compare our algorithm's interactivity and total isosurface computation times against
BCMC by comparing per-pass (\Cref{fig:eval_perf_scaling_xps13_pass,fig:eval_perf_scaling_m1mini_pass,fig:eval_perf_scaling_3080_pass})
and total times (\Cref{fig:eval_perf_scaling_xps13_total,fig:eval_perf_scaling_m1mini_total,fig:eval_perf_scaling_3080_total})
against the surface extraction times achieved by BCMC.
We find that our algorithm provides better interactivity
through its progressive rendering approach in all but two cases,
the Plasma on the XPS~13 and RTX~3080.
The interactivity improvement achieved by our method is especially pronounced on data sets with large and
complex isosurfaces such as the Miranda, where BCMC struggles with the large number of
active blocks and the size of the surface mesh.
At 1080p on the Miranda we achieve $7.2\times$, $9.5\times$ and $2.9\times$
faster pass times on XPS~13, M1 Mac Mini and RTX~3080 respectively, compared to BCMC's
surface extraction times.

Moreover, our algorithm achieves faster total surface computation
times than BCMC on the XPS~13 and M1 Mac Mini on the Miranda at all resolutions.
On the Miranda at 1080p we achieve speedups over BCMC of $1.4\times$ and $1.9\times$ on the
XPS~13 and M1 Mac Mini respectively.
On the Miranda at 720p these speedups grow to $2.2\times$ and $2.7\times$ on the
XPS~13 and M1 Mac Mini respectively.
On the Chameleon at 720p we achieve total surface computation times on
par with BCMC on the XPS~13 and M1 Mac Mini, while requiring substantially less memory.
BCMC typically outperforms our approach on the RTX~3080, though we do achieve
a $1.3\times$ speed-up at 360p on the Miranda.

\subsection{Memory Consumption}
\label{sec:eval_vs_bcmc}

\begin{table}[t]
    \centering
    \relsize{-1}{
    \begin{tabular}{@{}lrrr@{}}
        \toprule
        Data set & BCMC Avg. Mem & Our Avg. Mem & Reduction \\
        \midrule
        Skull & 329MB & 109MB & 3.02$\times$ \\
        TACC & 187MB & 108MB & 1.73$\times$  \\
        Plasma & 563MB & 191MB & 2.95$\times$ \\
        Kingsnake & 1.34GB & 607MB & 2.22$\times$ \\
        Chameleon & 2.09GB & 691MB & 3.02$\times$ \\
        Beechnut & --- & 1.06GB & --- \\
        Miranda & 4.20GB & 737MB & 5.70$\times$ \\
        JICF Q & --- & 1.00GB & --- \\
        DNS & --- & 875MB & --- \\
        R-M & --- & 4.19GB & --- \\
        \bottomrule
    \end{tabular}
    }
    \vspace{-0.5em}
    \caption{\label{tab:mem_vs_bcmc}%
    The average total compute memory overhead required by our algorithm vs. BCMC.
    We achieve an average $3.1\times$ reduction in total memory overhead.
    Entries marked by --- crashed due to exceeding the 4GB buffer binding limit in WebGPU.}
    \vspace{-1em}
\end{table}

\begin{table}[t]
    \centering
    \relsize{-1}{
    \begin{tabular}{@{}lrrr@{}}
        \toprule
        Data set & BCMC Avg. Cache Mem & Our Avg. Cache Mem & Reduction \\
        \midrule
        Skull & 66.0MB & 16.9MB & $3.9\times$ \\
        TACC & 55.0MB & 16.0MB & $3.4\times$  \\
        Plasma & 107MB & 44.7MB & $2.4\times$ \\
        Kingsnake & 545MB & 145MB & $3.8\times$ \\
        Chameleon & 375MB & 102MB & $3.7\times$ \\
        Beechnut & 1.99GB & 243MB & $8.2\times$ \\
        Miranda & 1.40GB & 167MB & $8.4\times$ \\
        JICF Q & --- & 170MB & --- \\
        DNS & 2.02GB & 406MB & $5\times$ \\
        R-M & --- & 522MB & --- \\
        \bottomrule
    \end{tabular}
    }
    \vspace{-0.5em}
    \caption{\label{tab:cache_mem_vs_bcmc}%
    The average cache size required by our algorithm vs BCMC.
    Our progressive wavefront traversal achieves a significant reduction in
    the volume working set size, providing a $4.8\times$ reduction in cache size on average.
    Entries marked by --- crashed due to the cache exceeding the 4GB buffer binding limit in WebGPU.}
    \vspace{-2em}
\end{table}

Finally, we compare the memory overhead of our technique against BCMC~\cite{usher_interactive_2020}.
BCMC provides a direct comparison point for explicit
isosurface extraction algorithms, as it also works directly on compressed data sets
and performs on the fly decompression to reduce memory overhead.
We report average memory statistics over the 100 random isovalue and 10 camera
position orbit benchmarks, rendering at $1280\times720$
(\Cref{tab:mem_vs_bcmc,tab:cache_mem_vs_bcmc}).

We achieve an average memory overhead reduction of $3.1\times$ 
compared to BCMC on the data sets BCMC is able to compute on without running out of memory
(\Cref{tab:mem_vs_bcmc}).
These memory reductions are achieved through our algorithm's
use of implicit ray-isosurface intersection, which eliminates the need for a
large vertex buffer, and our progressive wavefront traversal,
which significantly reduces the amount of data that must decompressed
to render the isosurface.

Furthermore, BCMC failed to compute the isosurface on the Beechnut, JICF Q, DNS, and R-M,
due to exceeding WebGPU's buffer size limit of 4GB.
These large data sets have noisy or turbulent
isosurfaces, resulting in some isosurfaces containing over 500M triangles.
Even with BCMC's quantized vertex format, these large isosurfaces exceed
4GB, resulting in a crash.
These results were run on the RTX~3080, which has 12GB of GPU memory;
however, on the XPS~13 or M1 Mac Mini these data sets would fail due to
running out of GPU memory, even if the size limit was lifted or otherwise worked around.
Our algorithm is able to achieve interactive rendering of these massive isosurfaces,
even on the XPS~13 and M1 Mac Mini.

\begin{figure}
    \centering
    \begin{subfigure}{\columnwidth}
        \centering
        \includegraphics[width=\textwidth]{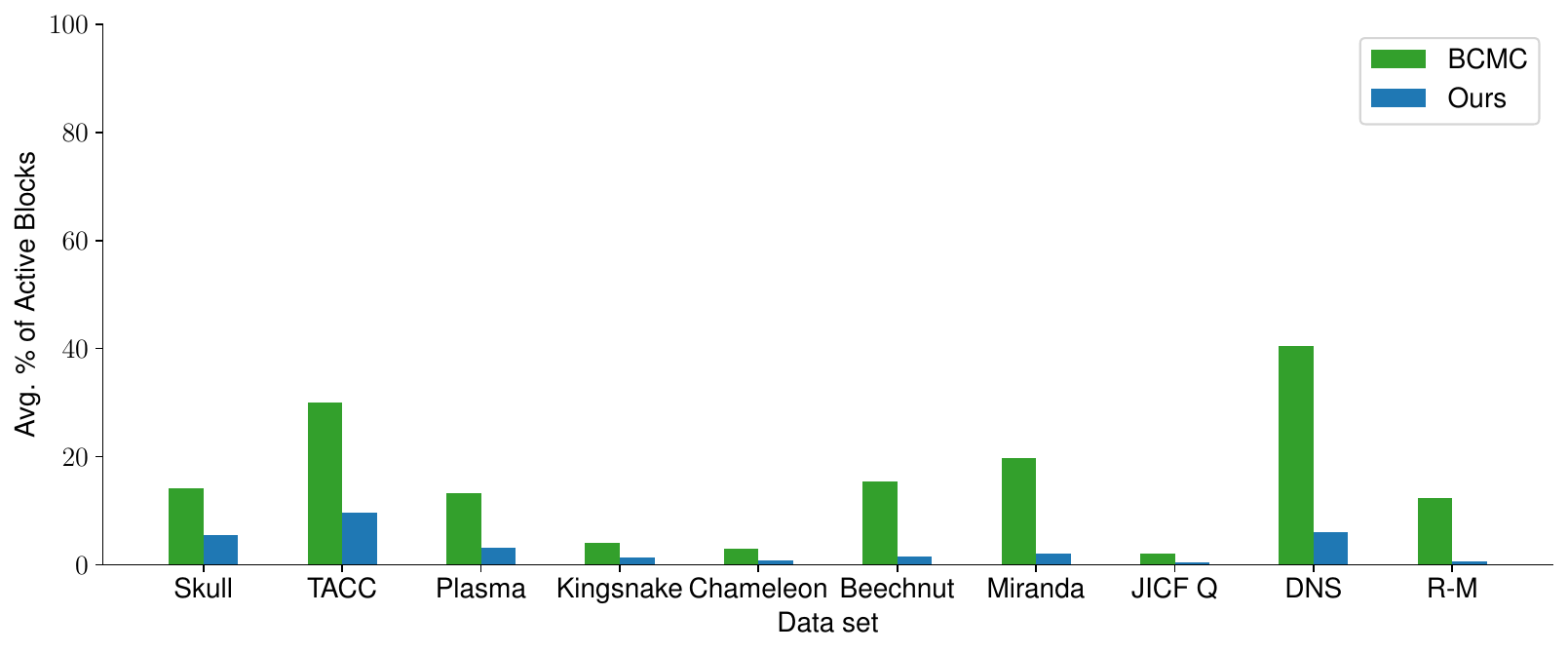}
        \vspace{-1.5em}
        \caption{\label{fig:active_block_percent_comparison_avg}%
        The average percentage of active blocks.}
    \end{subfigure}
    \begin{subfigure}{\columnwidth}
        \centering
        \includegraphics[width=\textwidth]{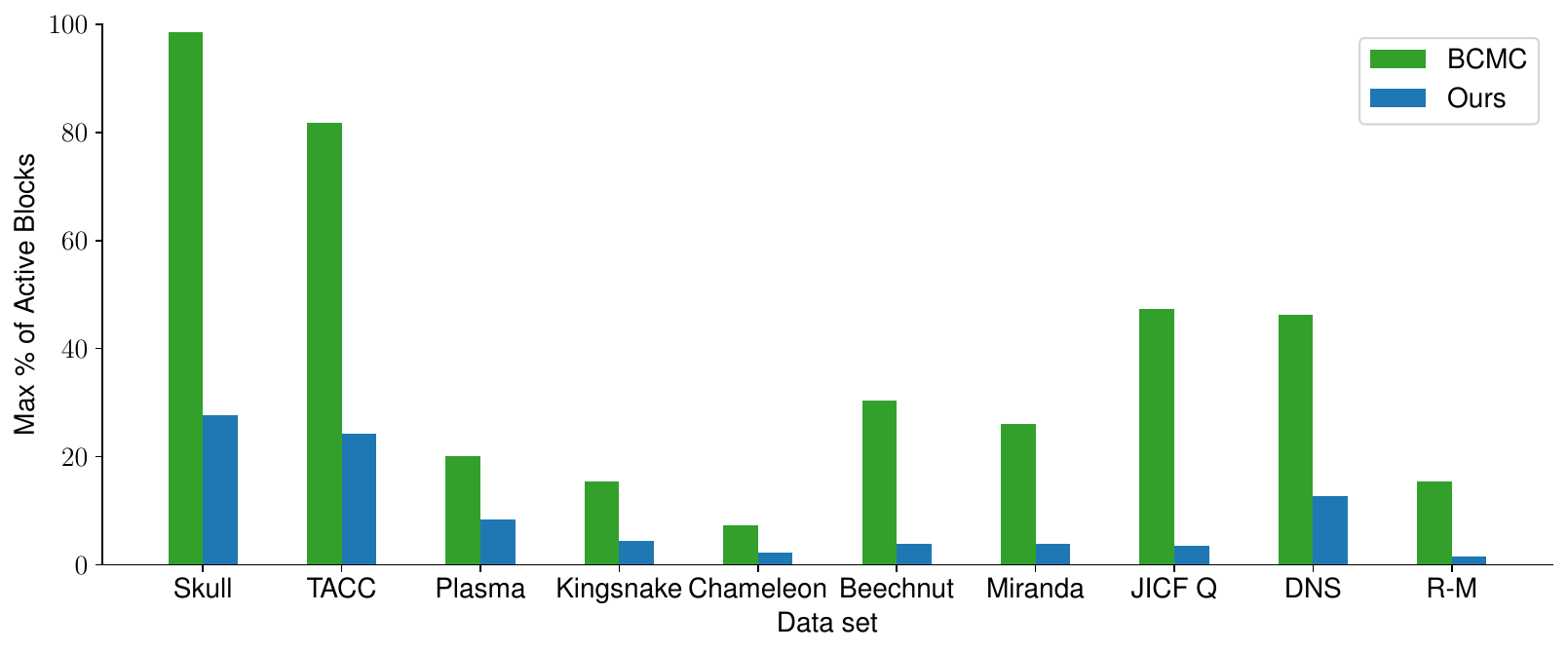}
        \vspace{-1.5em}
        \caption{\label{fig:active_block_percent_comparison_max}%
        The max percentage of active blocks.}
    \end{subfigure}
        \vspace{-1.75em}
    \caption{\label{fig:active_block_percent_comparisons}%
    The (a) average and (b) max percentage of active blocks required by BCMC and
    our algorithm. Our approach updates the cache each pass,
    storing just the blocks needed by active rays.
    In contrast, BCMC decompresses all blocks that may contain the isosurface.}
        \vspace{-1.5em}
\end{figure}

By progressively stepping rays through the volume and decompressing and caching just the blocks required
for each pass, we achieve significant reductions in the amount of data that must be decompressed.
\Cref{tab:cache_mem_vs_bcmc} compares the average cache memory required by BCMC
and our algorithm.
The Miranda and DNS results for BCMC were measured by disabling the vertex extraction step.
However, on the JICF Q and R-M, BCMC's active block cache memory alone exceeded 4GB, resulting in a crash.
We achieve an average cache size reduction of $4.8\times$ compared to BCMC on the data sets it is able to compute,
with far greater reductions achieved on the Beechnut ($8.2\times$) and Miranda ($8.4\times$).
The Beechnut is a noisy microCT scan and the Miranda is from a turbulent fluid mixing
simulation, resulting in large numbers of active but occluded blocks being decompressed and processed by BCMC.

Our view-dependent algorithm achieves substantial reductions in the number of blocks that must be decompressed
(\Cref{fig:active_block_percent_comparisons}).
These reductions come from a number of factors:
our algorithm can replace unneeded blocks with active ones each pass to minimize its working set;
blocks that are occluded or otherwise not visible are not decompressed;
and the number of visible blocks is driven by the image size and view position.
Compared to BCMC, we achieve a $6.7\times$ reduction in the average number of active blocks
and a $5.7\times$ reduction in the maximum number of active blocks.
The JICF~Q and R-M results for BCMC were measured by only recording the number of active blocks for
each isovalue and skipping all other computation to avoid crashing.




\section{Conclusion and Limitations}
\label{sec:conclusion}

We have proposed a new view-dependent isosurface rendering algorithm
designed specifically for interactive visualization of massive
isosurfaces on lightweight consumer platforms.
This is achieved through a progressive wavefront ray traversal algorithm
with per-pass block cache updates, where blocks of the data are decompressed and cached
on demand for each pass.
We accelerate isosurface rendering completion and increase GPU utilization
by introducing ray-block speculation into the algorithm.
Speculation enables us to fill open compute slots left by terminated rays
with speculated ray-block intersections for active rays to better leverage the GPU's
parallel compute power to complete the rendering in fewer passes and less time.

Our progressive, view-dependent isosurface rendering algorithm
is well suited to large scale isosurface visualization on end-user devices.
The compute and memory costs of our algorithm are not strongly affected by data set size
and can be easily reduced to scale it
down to lightweight systems and mobile devices by simply reducing the image resolution.
Furthermore, the progressive rendering provided by our algorithm makes it well suited
to provide low-latency interactive visualization.
Our algorithm runs entirely in the browser on the GPU through WebGPU
to expand access to large scale data visualization,
and is available on GitHub\footnote{https://github.com/Twinklebear/webgpu-prog-iso},
along with a live demo\footnote{http://progiso-demo.willusher.io}.

However, our approach is not without its limitations.
Although our method scales up well to large data sets, 
it does not scale down to small data sets.
For example, on the XPS13 and M1 Mac Mini BCMC achieves faster surface extraction
times on the Plasma and, in some cases, on the Chameleon.
Our approach still uses less memory on these data sets; however,
BCMC's overhead on smaller data sets is likely acceptable
for the performance improvement.
Further optimization efforts would be worthwhile to improve performance on
smaller data sets, improve scalability with image size,
and reduce overhead to improve per-pass and total rendering times overall.
We also find call overhead in JavaScript and WebGPU and note that better
performance could be achieved with a CUDA implementation where optimized
libraries such as Thrust and CUB are available. Bringing these libraries to WebGPU
would be a valuable effort.


There are also a number of interesting avenues left open for future work.
Although our speculation approach increases utilization and achieves large speed-ups in
total surface rendering time, our use of a global speculation count for all rays is restrictive.
It may be possible to achieve higher utilization by tracking a per-ray speculation
count; however, the added complexity may introduce additional overhead.
It would also be worthwhile to explore other acceleration structures that can be
built over the macrocell grid instead of our two-level grid to improve space skipping
and provide level of detail or multiresolution hierarchies to address current
limitations of our method with respect to undersampling the data.
For example, an implicit $k$-d tree~\cite{wald_faster_2005} built over the blocks
could further accelerate empty space skipping, or multiresolution and compression
techniques from work on compressed volume rendering could be integrated~\cite{balsa_rodriguez_state---art_2014-1,mensmann_gpu-supported_2010,suter_interactive_2011,wang_application-driven_2010,fout_high-quality_2005,schneider_compression_2003,fout_transform_2007,gobbetti_covra_2012}.
Leveraging multiresolution hierarchies within our method would address limitations
with respect to undersampling of the high-resolutio data, and enable rendering
larger data sets.
To improve image quality, it would be worth exploring support for secondary ray tracing effects in
our pipeline to add shadows, ambient occlusion, and global illumination with denoising.

Finally, as our algorithm's rendering and memory costs are primarily driven by the number of rays
traced and the number of passes, it would be worthwhile to combine it
with machine learning approaches for image up-scaling~\cite{weiss_volumetric_2021},
image in-painting, and foveated rendering~\cite{bauer_fovolnet_2023}
Such a combination would reduce the image resolution, number of passes, and rays traced
respectively; potentially reducing total surface rendering times to the cost of one or two passes
in our current method, within the same memory footprint.








\acknowledgments{This work was funded in part by NSF RII Track-4 award 2132013, NSF PPoSS planning award 2217036, NSF PPoSS large award 2316157 and, NSF collaborative research award 2221811.}

\bibliographystyle{abbrv-doi}
\bibliography{auto-exported,non-auto-bib}

\end{document}